\documentclass[twocolumn,pra,showpacs,amsmath,amssymb]{revtex4-1}
\usepackage[dvipdfmx]{graphicx}
\usepackage{setspace}
\usepackage{color}
\usepackage{ulem} 
\graphicspath{{./figures/}}

\usepackage{amsmath,amsthm,amssymb}
\usepackage{mathrsfs}
\usepackage{braket,comment}
\usepackage{bm}

\usepackage[colorlinks=true, citecolor=blue, urlcolor=blue, linkcolor=blue,
setpagesize=false, bookmarks=false
]{hyperref}

\begin{document}
\title{Role of Majorana fermions in spin transport of anisotropic Kitaev model}

\author{Hirokazu Taguchi}
\affiliation{Department of Physics, Tokyo Institute of Technology, Meguro, Tokyo 152-8551, Japan}
\author{Yuta Murakami}
\affiliation{Department of Physics, Tokyo Institute of Technology, Meguro, Tokyo 152-8551, Japan}
\author{Akihisa Koga}
\affiliation{Department of Physics, Tokyo Institute of Technology, Meguro, Tokyo 152-8551, Japan}
\author{Joji Nasu}
\affiliation{Department of Physics, Yokohama National University, Hodogaya, Yokohama 240-8501, Japan}
\affiliation{PRESTO, Japan Science and Technology Agency, Honcho Kawaguchi, Saitama 332-0012, Japan}
\date{\today}

\begin{abstract}
  We study a quantum spin Kitaev model with zigzag edges
  to clarify the effects of anisotropy in the exchange couplings on the spin propagation.
  We simulate the spin and Majorana dynamics triggered by a magnetic pulse,
  using the real-space time-dependent Majorana mean-field theory.
  When the anisotropy is small, the dispersion of the itinerant Majorana fermions remains gapless,
  where the velocity of the spin propagation matches the group velocity of
  the itinerant Majorana fermions at the nodal points.
  On the other hand, in the gapped system with a large anisotropy,
  the spin propagation is strongly suppressed
  although its nature depends on the shape of the pulse.
  The spin transport in the junction system described by the Kitaev models with distinct anisotropies
  is also addressed.
\end{abstract}

\maketitle

\section{Introduction}
Recently, spin transport has been attracting much interest.
One of the important mechanisms is the spin current induced by
a polarized electric current in metallic ferromagnets.
Such a spin current has intensively been studied
~\cite{PhysRevB.54.9353, PhysRevLett.85.5432, Spintronics, Tsoi2000,SpinCurrent-1, PhysRevLett.87.187202, PhysRevB.39.6995, Ogawa2016}.
Another is spin current in conventional insulating magnets,
where magnons carry spins without the electric current ~\cite{PhysRevLett.27.1729, PhysRevLett.74.3273, SpinCurrent-2, SpinCurrent-3}.
The common feature is that the spin current is realized in materials with magnetic orders.
On the other hand, it has been revealed that spin transport
is also realized in quantum spin liquids (QSLs), 
where long-range magnetic order is suppressed even at the zero temperature 
due to strong quantum fluctuations~\cite{ANDERSON1973153, Savary_2016, qsl1, qsl2, PhysRevB.88.041405, PhysRevB.40.7133, PhysRevB.44.2664, RevModPhys.89.025003}.
One of the typical examples is provided by an antiferromagnetic $S=1/2$ Heisenberg chain at the ground state.
The anisotropic negative spin Seebeck effect in the candidate material
$\text{Sr}_2\text{Cu}\text{O}_3$ 
indicates the spin current mediated by spinons~\cite{SpinonSpinCurrent},
which are magnetic elementary excitations in this system.
 
Another interesting playground for QSLs is given by the Kitaev model~\cite{Kitaev_model}.
The model is composed of bond-dependent Ising interactions between $S=1/2$ spins on the honeycomb lattice. 
One of the most remarkable features in this model is the existence of the local conserved quantity~\cite{Kitaev_model, PhysRevB.76.193101, PhysRevLett.98.087204, Kitaev_review, Motome_Nasu_Review}.
This guarantees the ground state to be a QSL
where the spin-spin correlation is exactly zero except for the nearest-neighbor sites.
The conserved quantity also leads to the existence of the spin fractionalization
and the spin degrees of freedom are split into the itinerant and localized Majorana fermions, 
the latter of which correspond to fluxes~\cite{Kitaev_review}.
Unlike spinons in the one-dimensional Heisenberg system, the Majorana fermions are not accompanied by the spin excitations naively, 
and hence it remains unclear whether they are capable of carrying the spin current.
To clarify this issue, the spin transport through QSL was studied in the isotropic Kitaev model in Refs.~\onlinecite{spin_trans,spin_trans_2}.
It was found that the spin excitation can propagate with a certain velocity
through the QSL regime without inducing spin polarizations.
The spin propagation turns out to be mediated by the itinerant Majorana fermions and the velocity of the spin propagation directly reflects the dispersion of the itinerant Majorana fermions.
The results suggest the close relationship between the spin transport through the Kitaev QSL and the  low-energy properties of the itinerant Majorana fermions, which can be controlled by changing the exchange couplings.
In this paper, we study the spin propagation in the anisotropic Kitaev model using real-space time-dependent Majorana mean-field theory ~\cite{Nasu_2018, MajoranaMF-1}.
The model exhibits the quantum phase transition between the gapless and gapped QSLs ~\cite{Kitaev_review}, and
we reveal the effects of anisotropy in the exchange couplings
on the spin propagation.

The paper is organized as follows.
In Sec.~\ref{sec:model},
we introduce the Kitaev model on the honeycomb lattice
and explain the Majorana mean-field theory.
In Sec.~\ref{sec:result},
we discuss how the anisotropy in the exchange interactions affects
the spin transport in the Kitaev model.
A summary is given in the last section.

\section{Model and method} \label{sec:model}

\begin{figure}[t]
  \centering
  \includegraphics[width=\linewidth]{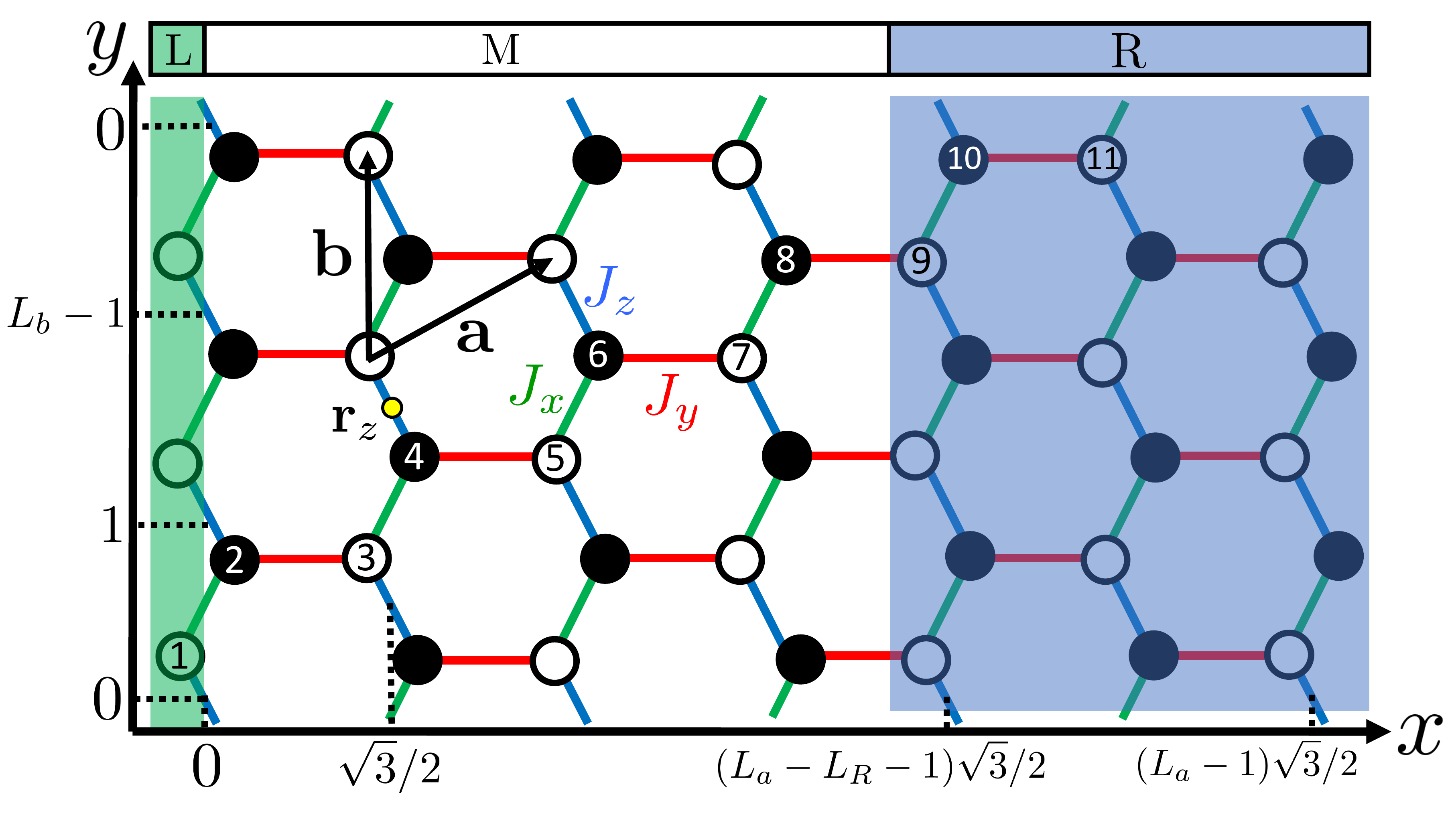} 
  \caption{The $L_a\times L_b$ cluster of the Kitaev model on the honeycomb lattice with zigzag edges. 
    $\bold{a}$ and $\bold{b}$ are primitive translational vectors.
    Green, red and blue lines indicate $x,y$, and $z$ bonds, respectively. 
    Solid (open) circles indicate spin-1/2 in the $A\ (B)$ sublattice 
    and the numbers in circles are the sequence of Jordan-Wigner transformation.
    In this figure, $L_a=7, L_b=3$, $L_R=3$ and
    its lattice constant is $1/\sqrt{3}$.
  }
  \label{fig:system}
\end{figure}

We study the spin transport through the QSL region in the Kitaev model on a two-dimensional honeycomb lattice.
To this end, we consider the Kitaev cluster shown in Fig.~\ref{fig:system}, 
where zigzag edges appear along a certain direction while
the periodic boundary condition is imposed in the other.
The system is composed of L, M, and R regions, where
the distinct magnetic fields are applied.
In the L region on the left edge, a time-dependent magnetic field $h_L(t)$ is applied.
No magnetic field is applied to the M region, while
the static magnetic field $h_R$ is applied to the R region.
The model Hamiltonian is given as 
\begin{eqnarray}
  H(t) &=& -\sum_{\mu = x,y,z}J_\mu \sum_{\braket{i,j}_\mu}S_i^\mu S_j^\mu \nonumber\\
  &&- h_R \sum_{i \in R}S_i^z - h_L(t)\sum_{i \in L}S_i^z,\label{eq:original}
\end{eqnarray}
where $\braket{i,j}_\mu$ indicates the nearest-neighbor sites on the $\mu(=x,y,z)$-bonds.
The $x$-, $y$-, and $z$-bonds are shown as green, red, and blue lines, respectively
in Fig.~\ref{fig:system}.
$S_i^\mu$  is the $\mu$ component of an $S=1/2$ spin operator at site $i$.
$J_\mu\;(\mu=x,y,z)$ is the exchange coupling on the $\mu$-bonds.

To discuss the real-time dynamics in the model \eqref{eq:original},
we represent the Hamiltonian in terms of Majorana fermions.
First, we regard the honeycomb lattice as a set of one-dimensional chains
composed of the $x$- and $y$-bonds, as shown in Fig.~\ref{fig:system}.
Then, the spin operators are described with the spinless fermions
using the Jordan-Wigner transformations as
\begin{align}
&S_i^+ = \prod_{j=1}^{i-1}(1-2a_j^\dagger a_j)a_i^\dagger, \\
&S_i^- = \prod_{j=1}^{i-1}(1-2a_j^\dagger a_j)a_i, \\
&S_i^z = a_i^\dagger a_i - \frac{1}{2},
\end{align}
where $a_i^\dagger$ and $a_i$ are the creation and annihilation operators
of the spinless fermion at the $i$th
site~\cite{PhysRevB.76.193101, PhysRevLett.98.087204, PhysRevLett.113.197205, Kitaev_review}.
We furthermore introduce two kinds of Majorana fermion operators for the $A\ (B)$ sublattice as 
\begin{align}
&\gamma_j^A = a_{j,A} + a_{j,A}^\dagger, \hspace{10pt} \bar{\gamma}_j^A = -i(a_{j,A} - a_{j,A}^\dagger), \\
&\gamma_j^B = -i(a_{j,B} - a_{j,B}^\dagger), \hspace{10pt} \bar{\gamma}_j^B = a_{j,B} + a_{j,B}^\dagger,
\end{align}
where $(\gamma_j)^\dag=\gamma_j$, $(\bar{\gamma}_j)^\dag=\bar{\gamma}_j$, $\gamma_j^2=\bar{\gamma}_j^2=1$, $\{ \gamma_i, \bar{\gamma}_j \} = 0$ ,
and $\{\gamma_i,\gamma_j\}= \{ \bar{\gamma_i}, \bar{\gamma}_j \} = 2\delta_{ij}$ ~\cite{Majorana2008, Wilczek2009}.
The Hamiltonian \eqref{eq:original} is rewritten as 
\begin{align}
H(t) = &-\frac{J_x}{4}\sum_{\bold{r}_z}i \gamma_{\bold{r}_z+\bold{b}}^A\gamma_{\bold{r}_z}^B 
- \frac{J_y}{4}\sum_{\bold{r}_z}i \gamma_{\bold{r}_z-\bold{a}+\bold{b}}^A\gamma_{\bold{r}_z}^B \notag \\
&- \frac{J_z}{4} \sum_{\bold{r}_z}i \gamma_{\bold{r}_z}^A\gamma_{\bold{r}_z}^B i\bar{\gamma}_{\bold{r}_z}^A\bar{\gamma}_{\bold{r}_z}^B \notag \\
&- \frac{h_R}{2} \sum_{\bold{r}_z \in R}(i \gamma_{\bold{r}_z}^A \bar{\gamma}_{\bold{r}_z}^A -  i \gamma_{\bold{r}_z}^B \bar{\gamma}_{\bold{r}_z}^B ) 
+ \frac{h_L(t)}{2} \sum_{\bold{r}_z \in L} i \gamma_{\bold{r}_z}^B \bar{\gamma}_{\bold{r}_z}^B, \label{H:jw}
\end{align}
where $\bold{r}_z$ is the position vector for the $z$-bond and
$\gamma^A_{\bold{r}_z}$ ($\gamma^B_{\bold{r}_z}$) is the Majorana fermion operator
at the $A \ (B)$ sublattice on the $z$-bond,
which is shown as the solid (open) circle in Fig.~\ref{fig:system}.

When $h_R=h_L(t)=0$, the operator
$\eta_{\bold{r}_z}= i \bar{\gamma}_{\bold{r}_z}^A\bar{\gamma}_{\bold{r}_z}^B$
commutes with the Hamiltonian and this model can be solved exactly,
where $\eta_{\bold{r}_z}$ is the $Z_2$ local conserved quantity.
This is because the second line of Eq.~\eqref{H:jw}, which originally describes the interaction between the two-types of Majorana fermions, is regarded as a one-body term.
On the other hand, $\eta_{\bold{r}_z}$ is no longer the conserved quantity in the regions under the magnetic field.
Then the model, in general, cannot be solved
since the magnetic fields induce the hybridization
between two types of the Majorana fermions, and thereby, the interaction between them is needed to be considered.
Here, we use the mean-field theory in the Majorana representation
given by Eq.~\eqref{H:jw}, and
the time-evolution is calculated within this formalism~\cite{Nasu_2018, MajoranaMF-1}.
Note that the cluster does not have the translational symmetry
in the $x$ direction perpendicular to the zigzag edges,
as shown in Fig.~\ref{fig:system}.
We introduce six kinds of time-dependent mean-field parameters, which are also functions of $x$, as 
\begin{align}
&\eta(x,t) = \braket{i\bar{\gamma}_{\bold{r}_z}^A \bar{\gamma}_{\bold{r}_z}^B}, \label{eq:mf1} \\
&\xi(x,t) = \braket{i\gamma_{\bold{r}_z}^A \gamma_{\bold{r}_z}^B}, \\
&m_A(x,t) = \frac{1}{2}\braket{i\gamma_{\bold{r}_z}^A \bar{\gamma}_{\bold{r}_z}^A}, \\
&m_B(x,t) = -\frac{1}{2}\braket{i \gamma_{\bold{r}_z}^B \bar{\gamma}_{\bold{r}_z}^B }, \\
&\Theta(x,t) = \braket{i \bar{\gamma}_{\bold{r}_z}^A \gamma_{\bold{r}_z}^B }, \\
&\Psi(x,t) = \braket{i\gamma_{\bold{r}_z}^A \bar{\gamma}_{\bold{r}_z}^B}, \label{eq:mf2}
\end{align}
where $\eta$ and $\xi$ are the expectation values of the localized and
itinerant Majorana fermions, and $m_\lambda(=\braket{S^z_\lambda})$ ($\lambda=A,B$) is the magnetization
at the $\lambda$-sublattice.  
Then, the interaction between the Majorana fermions given by the second line in Eq.~\eqref{H:jw} is decoupled
in terms of the Hartree-Fock approximation as 
\begin{align}
i &\gamma_{\bold{r}_z}^A\gamma_{\bold{r}_z}^B i\bar{\gamma}_{\bold{r}_z}^A\bar{\gamma}_{\bold{r}_z}^B \notag \\
\approx &i \gamma_{\bold{r}_z}^A \gamma_{\bold{r}_z}^B \eta(x,t)
+ \xi(x,t) i\bar{\gamma}_{\bold{r}_z}^A \bar{\gamma}_{\bold{r}_z}^B 
- \eta(x,t) \xi(x,t) \notag \\
&+ 2i \gamma_{\bold{r}_z}^A \bar{\gamma}_{\bold{r}_z}^A m_B(x,t)
- 2m_A(x,t) i \gamma_{\bold{r}_z}^B \bar{\gamma}_{\bold{r}_z}^B 
-4 m_A(x,t) m_B(x,t) \notag \\
&- i \gamma_{\bold{r}_z}^A \bar{\gamma}_{\bold{r}_z}^B \Theta(x,t)
- \Psi(x,t) i \bar{\gamma}_{\bold{r}_z}^A \gamma_{\bold{r}_z}^B 
+ \Theta(x,t)  \Psi(x,t). \label{eq:MF_int}
\end{align}
By solving the mean-field Hamiltonian self-consistently, we obtain the initial mean-field parameters and wave function.
After that, we evaluate the time-evolution of the ground state using extended Euler methods 
\cite{extend_euler-1,extend_euler-2,extend_euler-3,extend_euler-4,extend_euler-5,extend_euler-6}.
Since the mean-field theory gives the exact results
for $h_L(t) = h_R = 0$,
we believe that the obtained results are reliable as far as
the applied fields are small enough.
Details of the implementation of the Majorana mean-field theory and time-evolution is given in Appendix.~\ref{sec:imp_MF}.

    \begin{figure*}[t]
      \centering
      \includegraphics[width=\linewidth]{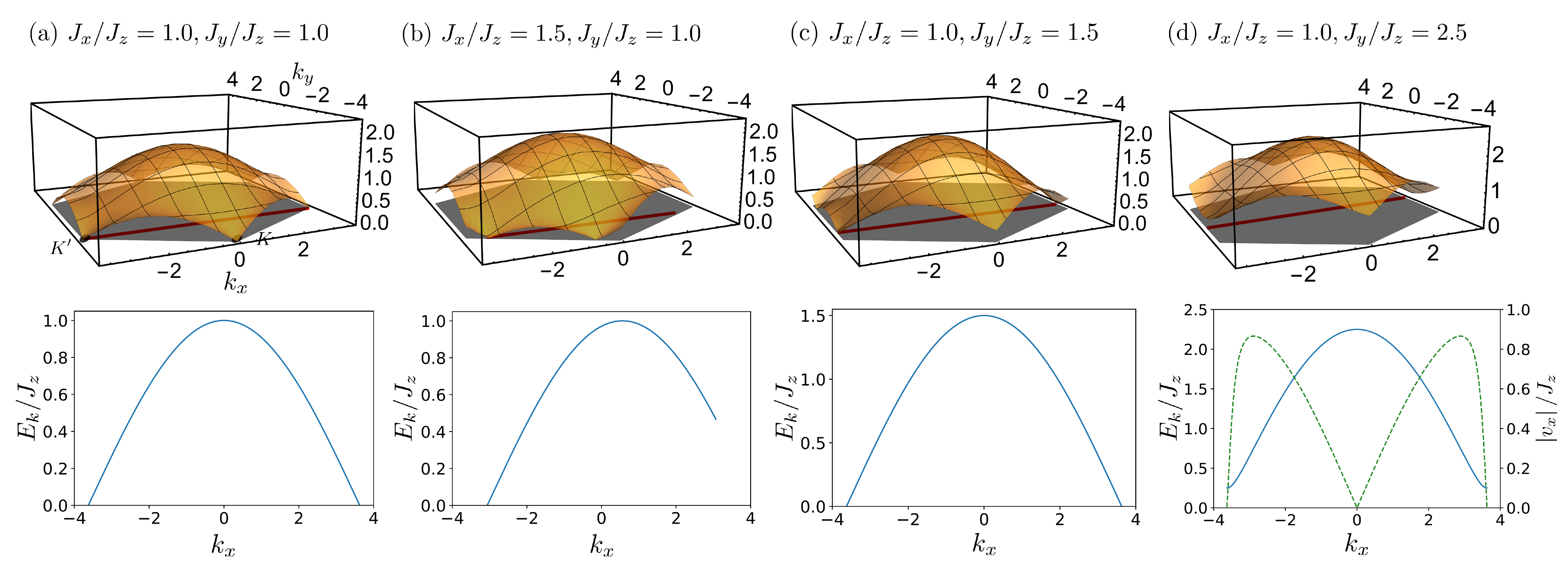} 
      \caption{
        Upper panels represent the dispersion relations of the itinerant Majorana fermions for parameters indicated. 
        Lower panels depict the dispersion relations along the red lines
        shown in the upper panels.
        The dashed line in (d) represents the Majorana velocity.
        }
  \label{fig:dispersion}
\end{figure*}
Before discussing the real-time dynamics,
we briefly review the dispersion relation of the 
itinerant Majorana fermions in the Kitaev model without the magnetic field,
which is closely related to the spin transport.
When the periodic boundary conditions are imposed in $x$ and $y$ directions,
the dispersion relation of the itinerant Majorana fermions $E(\bold{k})$
is obtained as~\cite{Kitaev_model}
\begin{align}
  E({\bf k})=\frac{1}{2}\left|
  J_x e^{i{\bf k}\cdot{\bf b}}+J_y e^{i{\bf k}\cdot({\bf b}-{\bf a})}+J_z
  \right|,
\end{align}
where ${\bf a}$ and ${\bf b}$ are the primitive lattice vectors shown in Fig.~\ref{fig:system}.
Here, we focus on the velocity defined by
\begin{eqnarray}
  {\bf v}(\bf{k}) &=& \nabla_{\bf k}E({\bf k}).\label{eq:velocity}
\end{eqnarray}
In the isotropic case ($J_x = J_y = J_z$),
gapless linear dispersions appear at ${\bf k}_0=K$ and $K'$ points
in the Brillouin zone,
where ${\bf k}_0$ is defined so that $E({\bf k}_0)$ takes a minimum.
Its velocity is given by
$\left| {\bf v}({\bf k}_0) \right| = (\sqrt{3}/4)J_z$ and does not depend on the direction around ${ \bf k}_0$, and
the low-energy dispersion can be regarded as an isotropic cone.
Beyond the isotropic case, the gapless dispersion appears as far as
the following inequalities are satisfied as
\begin{align}
&|J_x| + |J_y| \geq |J_z|, \label{eq:gapless_condition1}\\
&|J_y| + |J_z| \geq |J_x|, \\  
&|J_z| + |J_x| \geq |J_y|. \label{eq:gapless_condition3}
\end{align}
Figures~\ref{fig:dispersion}(b) and \ref{fig:dispersion}(c) show the dispersions for the Kitaev models
with the small anisotropy in the exchange couplings. 
It is found that the gapless points ${\bf k}={\bf k}_0$ are shifted
from the $K$ or $K'$ points.
The dispersion in the gapless state is then expanded
around ${\bf k}= {\bf k}_0$ as 
\begin{align}
&[E(\bold{k})]^2 \notag \\
&\sim \frac{3}{16}J_y^2 \tilde{k}^2_x + \frac{1}{16}\left[ 4J_x^2 + J_y^2 + 4J_xJ_y\cos{(\bold{k}_0 \cdot \bold{a})} \right] \tilde{k}_y^2 \notag \\
&+ \frac{\sqrt{3}}{8}(J_x^2 - J_z^2)\tilde{k}_x \tilde{k}_y, \label{eq:low_energy}
\end{align}
where $\tilde{\bold{k}} = (\tilde{k}_x, \tilde{k}_y)\equiv {\bf k}-{\bf k}_0$.
The velocity of the itinerant Majorana fermions ${\bf v}( {\bf k}_0) = (v_x({\bf k}_0), v_y({\bf k}_0))$ 
depends on both the direction in the $k$ space and the anisotropy in the exchange couplings. 
In the case with $J_x=J_z$ and $J_y<2J_z$,
the system is in the gapless state and
$v_x( {\bf k}_0)=(\sqrt{3}/4)J_y$.
\begin{figure}[htb]
  \centering
  \includegraphics[width=\linewidth]{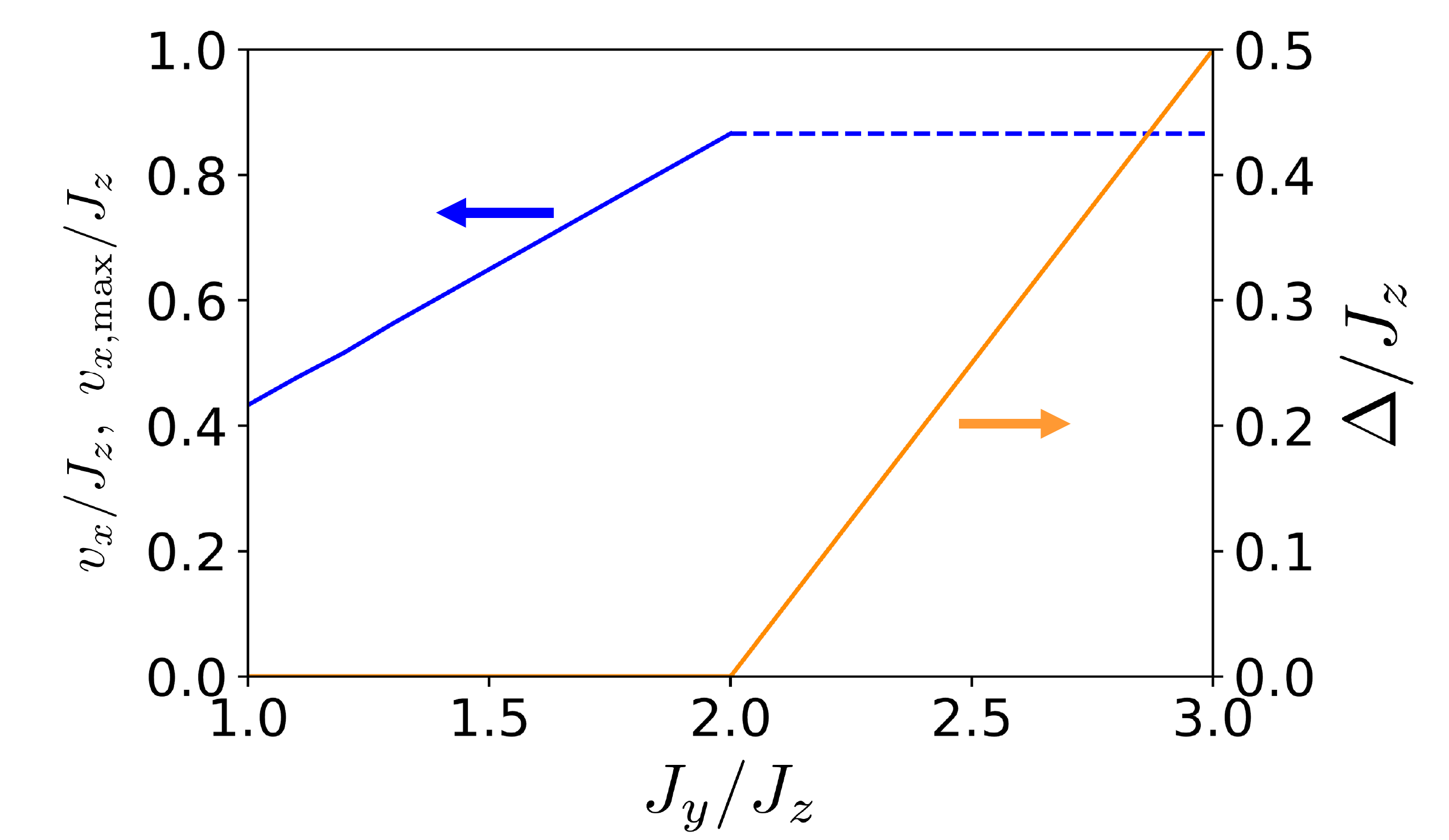} 
  \caption{
    The excitation gap $\Delta$ (orange line), and velocities $v_x$ (solid blue line) and $v_{x,\text{max}}$ (dashed blue line)
    as a function of $J_y/J_z$ in the Kitaev system on the honeycomb sheet with fixed $J_x/J_z$ = 1.0.
  }
  \label{fig:gap_velocity}
\end{figure}
On the other hand, when $J_y>2J_z$, the system has the excitation gap
in the Majorana dispersion, $\Delta\propto J_y-2J_z$, as shown in Fig.~\ref{fig:gap_velocity}.
In this state, $v_x$ is zero because of the quadratic band dispersion around ${\bf k}_0$. 
Instead, we numerically examine $v_{x, \text{max}}$, which is the maximum value of $v_x ( {\bf k})$ in the Brillouin zone.
In the case with $J_x/J_z=1.0$ and $J_y/J_z=2.5$,
$v_x({\bf k})$ is maximum at $k_x\sim \pm 2.9$ and $v_{x,\text{max}}\sim 0.87 \simeq	\sqrt{3}/2$,
in Fig.~\ref{fig:dispersion}(d).
The maximum values as a function of $J_y/J_z$
are shown as the dashed lines in Fig.~\ref{fig:gap_velocity}.
We find that $v_{x,\text{max}}$ is not changed in the gapped state and
coincides with $v_x({\bf k}_0)$ at the critical point $(J_y/J_z)_c$.
In the next section, we discuss the role of these velocities
for the spin transport in the gapless and gapped Kitaev systems.

In the present study, 
we consider the honeycomb lattice with $L_a = 200, L_b=300$, and $L_R = 50$.
Then we examine real-time dynamics in the Kitaev model
with anisotropic exchange couplings.
The static magnetic field $h_R$ in the R region is set to be $0.01J$,
which is smaller than the critical values
$h_c$~\cite{Nasu_2018,PhysRevB.98.054433,PhysRevB.101.045121}.
We introduce a Gaussian magnetic pulse as the time-dependent field in the L region,
which is given as
\begin{align}
h_L(t) = \frac{A}{\sqrt{2 \pi}\sigma} \exp{\left[ - \frac{t^2}{2\sigma^2} \right]},
\end{align}
where $A$ and $\sigma$ are strength and width of the pulse.
In the following, the width of the pulse
is mainly used as $\sigma=5.0/J_z$ and $A=1.0$.
Then, we study how the anisotropy in the exchange couplings affects
the spin transport in the Kitaev model.

\section{Results}\label{sec:result}

\begin{figure}[t]
  \centering
   \includegraphics[width=\linewidth]{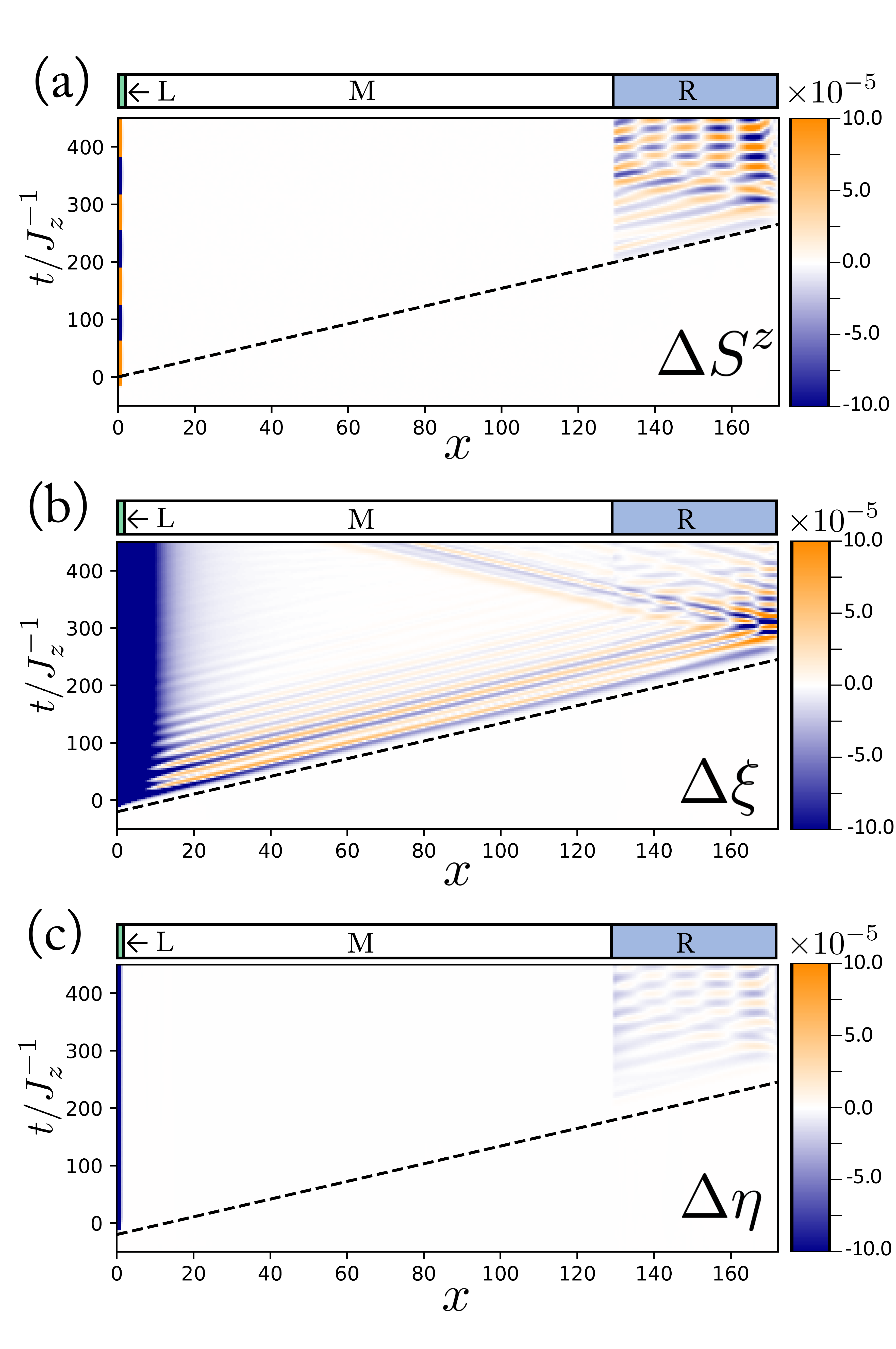} 
   \caption{Real-time evolution of (a) $\Delta S^z$, (b) $\Delta \xi$, and (c) $\Delta \eta$
     in the Kitaev system with $J_x/J_z = 1.0$ and $J_y/J_z= 1.5$.  
     Here, we use $A=1.0$ and $\sigma=5.0/J_z$ for the magnetic field pulse.
     The dashed lines represent $x = (\sqrt{3}/4) J_y t$ (see text). }
  \label{fig:result_101510}
\end{figure}

First, we focus on the Kitaev model with gapless dispersions
to discuss the spin propagation.
Figure~\ref{fig:result_101510} shows the change in the spin moment and
Majorana mean-fields
$\Delta S^z(x,t)$, $\Delta \xi(x,t)$, and $\Delta\eta(x,t)$
for the system with $J_x/J_z = 1.0$ and $J_y/J_z = 1.5$,
where $\Delta O(x,t)=O(x,t)-O(x,-\infty)$.
The ground state of the Kitaev model
without the external magnetic field
is the QSL, where the magnetic moment never appears~\cite{short_interaction}.
In fact, no magnetic moments are induced in the M region
even after the magnetic pulse is introduced in the L region,
as shown in Fig.~\ref{fig:result_101510}(a).
By contrast, finite oscillations in $\Delta S^z(x,t)$ emerges in the R region
after some time interval.
It is also found that the mean field for the itinerant Majorana fermions oscillates
in the whole region, while that for the localized Majorana fermions changes
only in the R region, as shown in Figs.~\ref{fig:result_101510}(b) and (c).
This means that the spin excitations are carried
by the itinerant Majorana fermions,
which are induced by the spin fractionalization in the Kitaev model.
Thus, we expect that the velocity of spin propagation is determined by that of
the itinerant Majorana fermions.
The dashed lines in Fig.~\ref{fig:result_101510} stand for the velocity of the itinerant Majorana fermions along the $x$ direction, $v_x({\bf k}_0)$.
We find that the emergence of the magnetic oscillation in the R region is well scaled by the motion of the Majorana fermions, 
implying that the change of the magnetization is induced by low-energy Majorana fermions.
Similar behavior is also observed in the case with $J_x/J_z=1.5$ and $J_y/J_z=1.0$ (not shown).
Thus, we confirm that the itinerant Majorana fermions around the gapless points
play an essential role for the spin transport in the gapless Kitaev model.

\begin{figure}[htb]
  \centering
   \includegraphics[width=\linewidth]{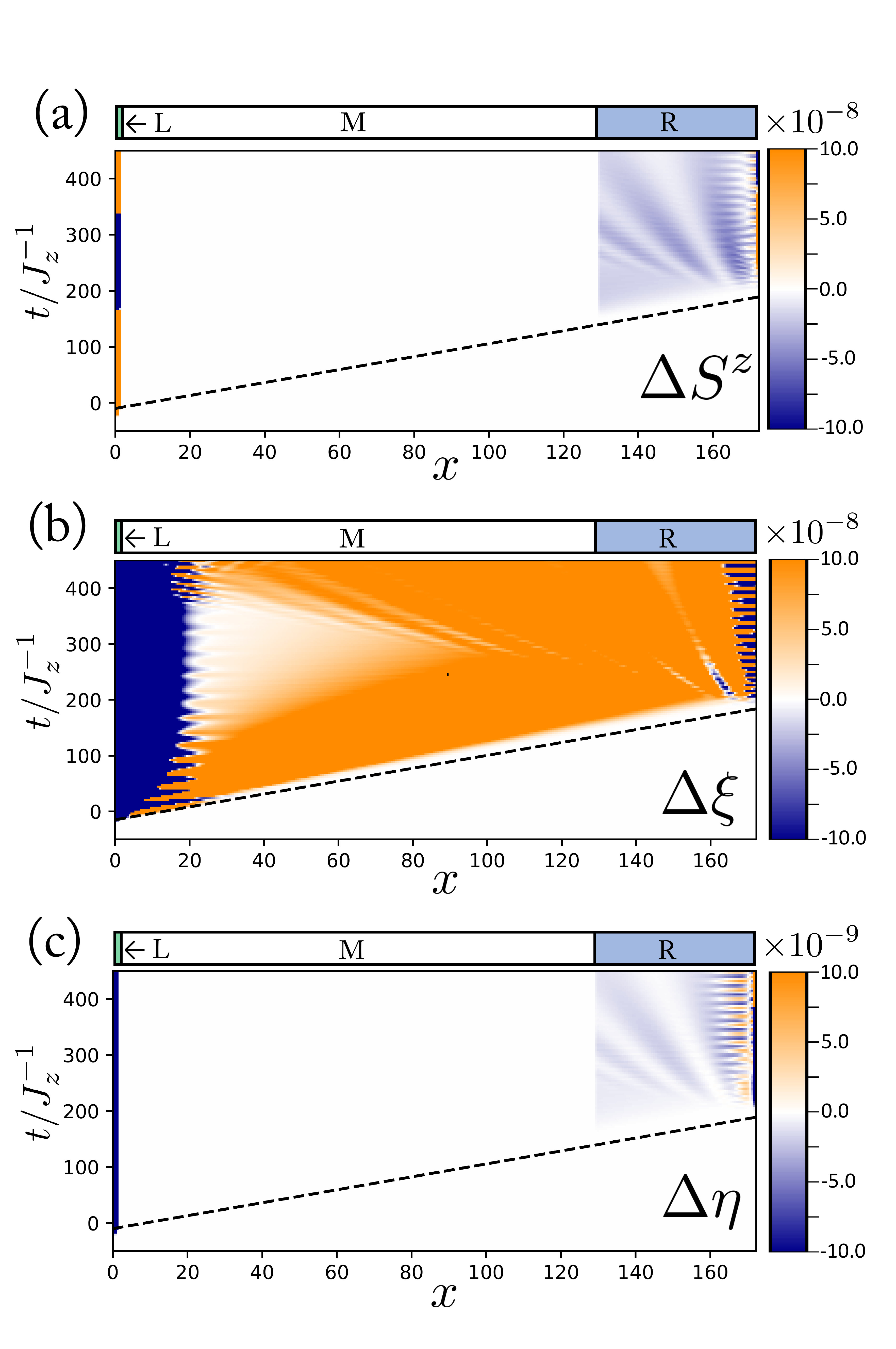} 
   \caption{Real-time evolution of (a) $\Delta S^z$, (b) $\Delta \xi$, and (c) $\Delta \eta$
     in the Kitaev system with $J_x/J_z=1.0$ and $J_y/J_z=2.5$.
     Here, we use $A=1.0$ and $\sigma=5.0/J_z$ for the magnetic field pulse.
     The dashed lines represent $x= v_{x, \text{max}} t$ (see text).
}
  \label{fig:result_102510}
\end{figure}
Next, we consider the Kitaev model with the large anisotropy in the exchange couplings 
to discuss the spin propagation in the gapped system.
When $J_x/J_z=1.0$ and $J_y/J_z=2.5$,
the system has the excitation gap $\Delta=0.25 J_z$.
Nevertheless, a similar spin propagation is observed
although its amplitude is much smaller. 
Figure~\ref{fig:result_102510} shows the time and space dependence of the mean fields for the above exchange parameters.
The spin moments never appear in the M region as presented in Fig.~\ref{fig:result_102510}(a) 
but small oscillations are induced in the R region
after some time interval.
The oscillation in the itinerant Majorana fermions appears in the whole region,
as shown in Fig.~\ref{fig:result_102510}(b).
This implies that the Majorana-mediated spin transport
occurs even in the gapped system
although spin and Majorana fluctuations are strongly suppressed due to the presence of the excitation gap.
Now, we focus on the velocity of the spin propagation.
The oscillation of mean fields propagates
with a certain velocity comparable to but a bit smaller than $v_{x,\text{max}}$,
which is shown as dashed lines in Fig.~\ref{fig:result_102510}.
This is due to the existence of the gap in the itinerant Majorana dispersion.
In the case with $\sigma=5.0/J_z$,
the pulse does not dominantly contribute to the Majorana fermions
with $v_{x, \text{max}}$ while it does to the lower-energy Majorana fermion, leading to the slightly slower spin propagation visible
in Fig.~\ref{fig:result_102510}.

Generally, the Gaussian pulse 
can be represented by the superposition of plane waves with distinct energies.
Thus, the magnetic field pulse excites itinerant Majorana fermions 
in a certain energy range $\lesssim \sigma^{-1}$.
To clarify the pulse dependence of the spin propagation in the gapped Kitaev system,
we focus on the itinerant Majorana fermions, which play an essential role for
the spin transport.
\begin{figure}[t]
  \centering
   \includegraphics[width=\linewidth]{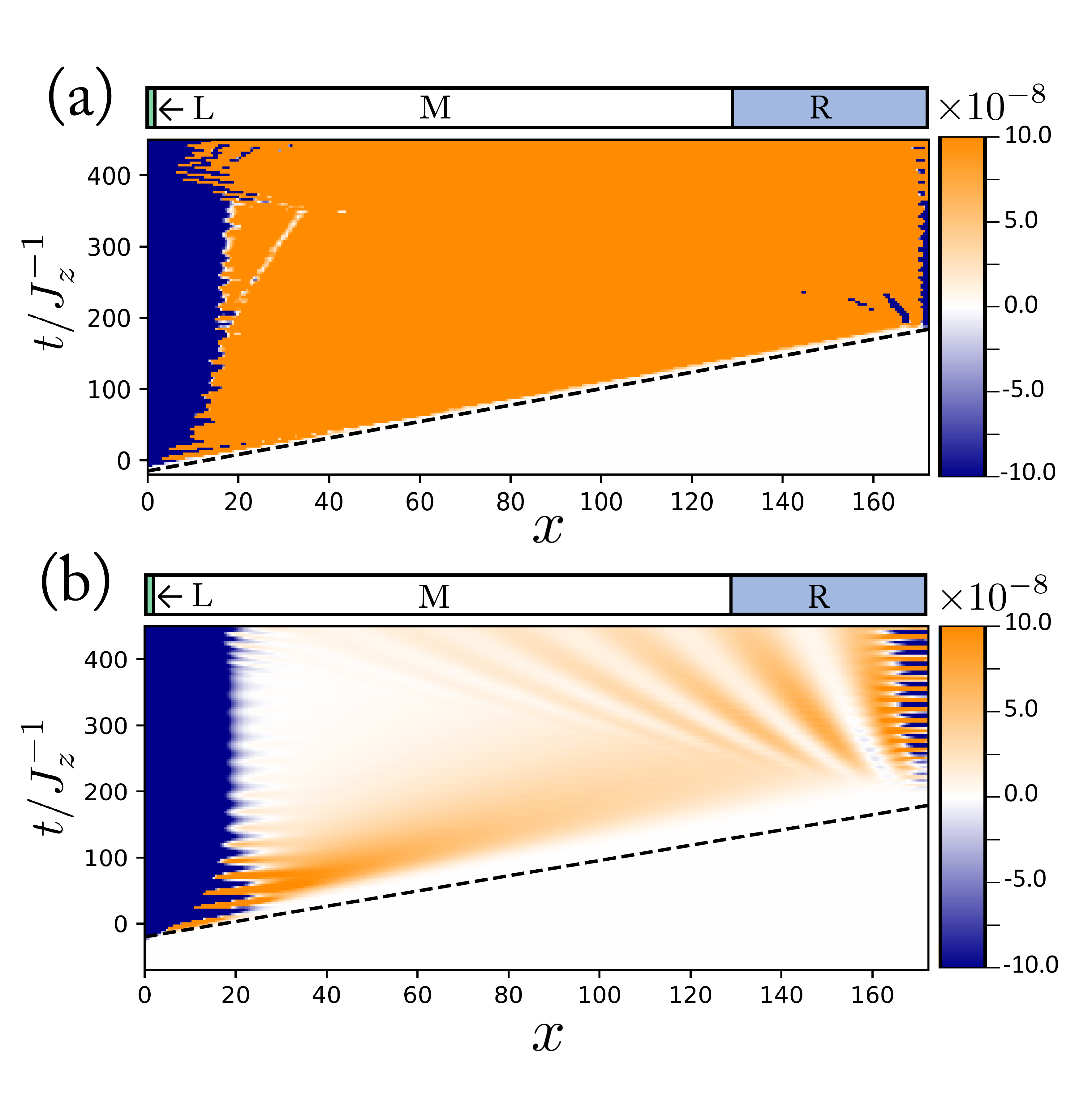} 
   \caption{Real-time evolution of $\Delta \xi$
     when the magnetic field pulses with (a) $\sigma = 2.0/J_z$ and (b) $\sigma = 7.0/J_z$
     are introduced in the gapped Kitaev system with $J_x/J_z=1.0$ and $J_y/J_z=2.5$. 
     The dashed lines represent $x = v_{x,\text{max}} t$.
     For comparison, we have the same scale of color map as that in Fig.~\ref{fig:result_102510}(b). 
     }
  \label{fig:result_gapful}
\end{figure}
Figure~\ref{fig:result_gapful} shows the change of the mean field, $\Delta\xi$,
when the magnetic field pulses with $\sigma=2.0/J_z$ and $7.0/J_z$ are injected.
It is clearly found that, in the case with a sharper pulse, 
the velocity of the spin transport corresponds to $v_{x,\text{max}}\sim 0.87$, which
is shown as the dashed line in Fig.~\ref{fig:result_gapful}(a).
By contrast, in the case with $\sigma=7.0/J_z$,
one finds that the oscillation propagates slowly.
Therefore, we can say that the spin transport in the gapped Kitaev model
depends on the form of the injected magnetic field.
These results are in contrast to those for the gapless state,
where low energy massless excitations always
play an essential role for the spin transport and
the change in $\sigma$ has little effect on its velocity.

Before conclusion, we consider the junction system composed of two Kitaev models
with distinct coupling constants. 
We discuss the effect of an interface on the Majorana excitations
triggered by the magnetic pulse.
\begin{figure}[htb]
  \centering
   \includegraphics[width=\linewidth]{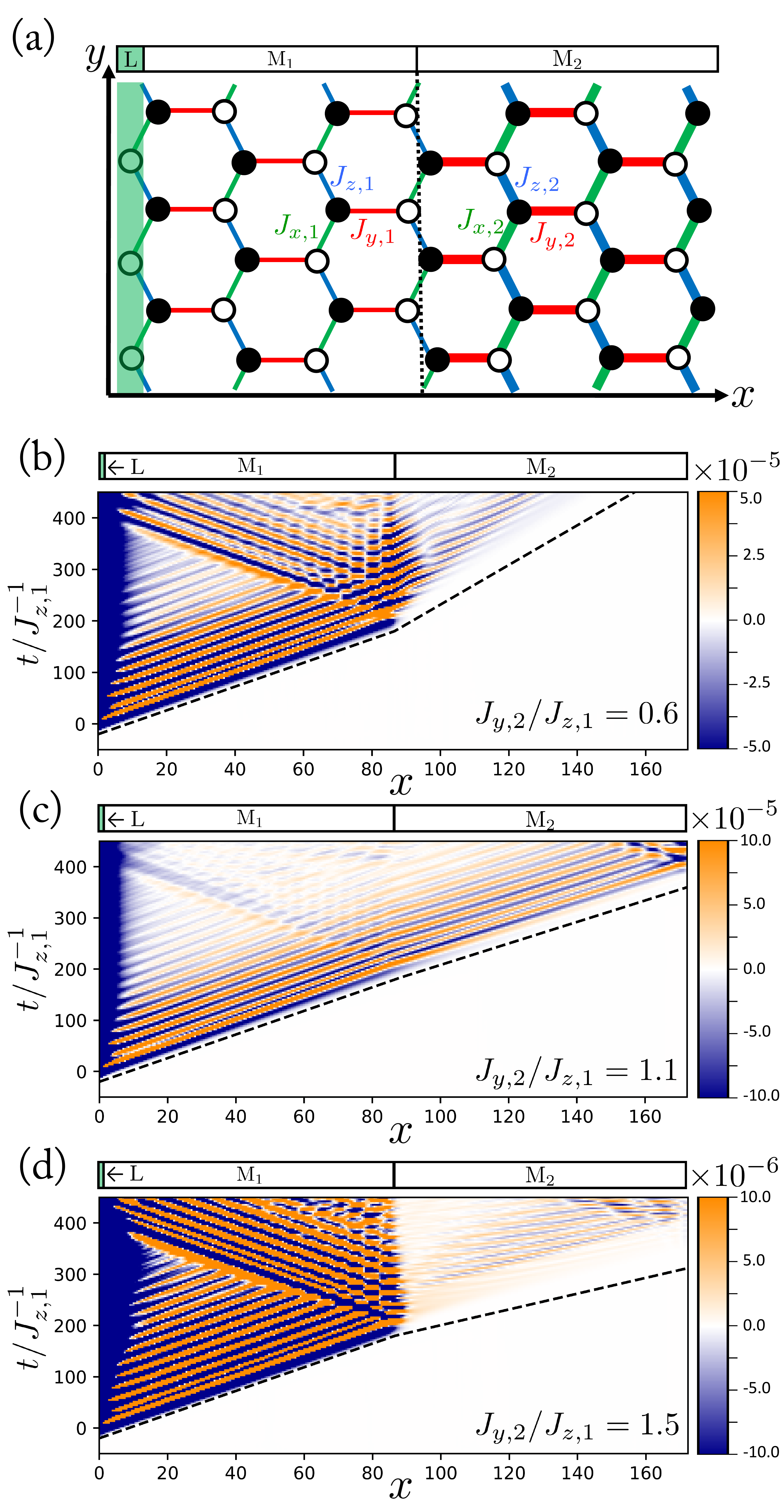} 
   \caption{
     (a) The zigzag-edge cluster composed of two Kitaev models.
     The interface of the junction is located at the center.
     Real-time evolution of $\Delta \xi$ in the Kitaev system with
     (b) $J_{y,2}/J_{z,1}=0.6$, (c) $1.1$, and (d) $1.5$.
     Each dashed line represents the Majorana velocity in the corresponding region.
}
  \label{fig:refraction}
\end{figure}
The cluster we treat here is composed of two regions $\text{M}_1$ and $\text{M}_2$ without static magnetic fields,
as shown in Fig.~\ref{fig:refraction}(a).
In the left region $\text{M}_1$, the system is the isotropic Kitaev model with $J_{x,1}=J_{y,1}=J_{z,1}$.
The right region $\text{M}_2$ is described by the anisotropic Kitaev model with $J_{x,2}=J_{z,2}=J_{z,1}$ and
$J_{y,2}\neq J_{z,1}$.
Then, the interface is located between two regions.
Here, we calculate the change in the mean field for the itinerant Majorana
fermions since it plays an important role for the spin transport as discussed above.
Figures~\ref{fig:refraction}(b)-~\ref{fig:refraction}(d) show $\Delta\xi$ in the systems with
$J_{y,2}/J_{z,1} = 0.6, 1.1$ and $1.5$.
Since the velocity of the itinerant Majorana fermions is suddenly changed at the interface,
the refraction occurs, yielding reflected and transmitted waves.
By introducing the anisotropy in $J_{y,2}$,
the Majorana oscillation smears in the right region.
The results indicate that
the reflection ratio increases associated with the decrease of the transmission ratio.
This is consistent with the conventional Fresnel's theorem, which says that the refraction ratio given by proportion of two kinds of velocities in $\text{M}_1$ and $\text{M}_2$.
In addition, low-energy properties such as the position of the nodal point in the momentum space are suddenly changed at the interface.
This should lead to a certain delay in the propagation at the interface.

Finally, we note that the mean-field analysis of the Kitaev model does not include 
effects of scattering of Majorana fermions.
Furthermore, for candidate materials
~\cite{RevModPhys.87.1, KitaevMaterial-1, KitaevMaterial-2, KitaevMaterial-3, KitaevMaterial-4, KitaevMaterial-5, KitaevMaterial-6}, 
the effects of additional terms beyond the Kitaev model as well as scattering with impurities should be considered.
With these effects, we expect that the spin transport immediately vanishes in the gapped case.
On the other hand, since low-lying itinerant Majorana fermions does not induce magnetic excitations in the bulk, 
long-range spin transport is expected to be retained even in the presence of magnetic impurities.
It is also interesting to examine Majorana correlations
in the present system~\cite{KogaArxiv},
which is beyond the scope of our study.

\section{Conclusion}\label{sec:conclude}
We have studied the spin transport in the Kitaev model with anisotropic exchange couplings
by means of the time-dependent Majorana mean-field theory.
When the anisotropy is small,
the dispersion of the itinerant Majorana fermions remains gapless.
The group velocity of the low-energy Majorana fermions along the $x$ direction 
is proportional to $J_y$, and it determines the velocity of the spin transport.
In the gapless cases, the spin transport is mediated by the itinerant Majorana fermions around the nodal points, and
hence spin excitations can travel over long distance regardless of the shape of magnetic field pulse.
When the anisotropy is large, the Majorana dispersion is gapped.
While the magnitude of spin oscillations is drastically reduced
in comparison with the gapless case,
the velocity of the spin propagation correlates with the group velocity
of the Majorana fermions above the gap.
However, we find that the difference between them 
is more apparent in the case with the wider magnetic field pulse.
We have also studied the junction of the Kitaev systems with different anisotropies of the exchange constants to show the reflection and transition of the itinerant Majorana fermions at the interface.
Since the manipulation of an anisotropy in the exchange coupling
was recently proposed in the realistic materials by means of the light irradiation~\cite{Arakawa},
the junction system would be a promising candidate for spintronic devices mediated by Majorana fermions.

\acknowledgements
We would like to thank T. Minakawa for valuable discussions.
Parts of the numerical calculations are performed in the
supercomputing systems in ISSP, the University of Tokyo. 
This work is supported by Grant-in-Aid for Scientific Research from JSPS, KAKENHI
Grant Nos. JP19K23425, JP20K14412, JP20H05265 (Y.M.),
JP21H01025, JP19H05821, JP18K04678, JP17K05536 (A.K.),
JP19K03742, JP20H00122, JST PREST Grant No. JPMJPR19L5 (J.N.),
and JST CREST Grant No. JPMJCR1901 (Y.M.).

\appendix

\section{Implementation of Majorana mean-field theory}\label{sec:imp_MF}

The Majorana mean-field theory is implemented for the present problem in the following way.
First, the Majorana mean-field Hamiltonian is obtained from Eq.~\eqref{H:jw} using Eqs.~\eqref{eq:mf1}-\eqref{eq:mf2}.
Since the system 
has a translational invariance along the ${\bf b}$-direction, we make the partial Fourier transformation
for $\gamma$ and $\bar{\gamma}$ ,
and express the mean-field Hamiltonian with them. Namely, we introduce
\begin{align}
  &c_{x,k}^A = \frac{1}{\sqrt{2L_b}} \sum_y \gamma^A_{\bold{r}_z} e^{-iky}, \\
  &\bar{c}_{x,k}^A = \frac{1}{\sqrt{2L_b}} \sum_y \bar{\gamma}^A_{\bold{r}_z} e^{-iky}, \\
  &c_{x,k}^B = \frac{1}{\sqrt{2L_b}} \sum_y \gamma^B_{\bold{r}_z} e^{-iky}, \\
  &\bar{c}_{x,k}^B = \frac{1}{\sqrt{2L_b}} \sum_y \bar{\gamma}^B_{\bold{r}_z} e^{-iky}, 
\end{align}
where ${\bf r}_z=(x,y)$ is the position vector for the $z$-bond, and we define the wave vector $k$ as only for $k > 0$.
In this case, the operators $c_{x,k}^{A(B)}$ and $\bar{c}_{x,k}^{A(B)}$ satisfy $c_{x, -k}^{A(B)} = \left( c_{x,k}^{A(B)} \right)^\dagger$ 
and the usual fermionic anticommutation relation as 
\begin{align}
  &\left\{ c_{x,k}^{A(B)} , \left( c_{x', k'}^{A(B)} \right)^\dagger \right\} = \delta_{x,x'} \delta_{k, k'}, \notag \\
  &\left\{ c_{x,k}^{A(B)} , c_{x', k'}^{A(B)} \right\} = 0, \notag \\
  &\left\{ \left( c_{x,k}^{A(B)} \right)^\dagger , \left( c_{x', k'}^{A(B)} \right)^\dagger \right\} = 0 \hspace{10pt} (k, k' > 0).
\end{align}
With these operators, the mean-field Hamiltonian can be expressed as
\begin{align}
  H^{\rm \text{MF} }(t) = \sum_{k>0} \Phi_k^\dagger H_k^{\text{MF}}(t,\{ \Theta_i (t) \} )\Phi_k,
\end{align}
where $\Phi_k = \{ c_{0,k}^A, c_{0,k}^B, \bar{c}_{0,k}^A, \bar{c}_{0,k}^B, c_{\sqrt{3}/2,k}^A, \cdots ,\bar{c}_{\sqrt{3}(L_a-1)/2,k}^B \}^T$
and $\Theta_i(t)\ (i=1,2,\ldots, 6)$ are six mean-field parameters.
Since $H_k^{\text{MF}}$ are $4L_a \times 4L_a$ Hermitian matrices, they can be diagonalized by unitary matrices $U_k(t, \{ \Theta_i (t) \})$ as 
\begin{align}
  &H_k^{\text{MF}}(t) = \sum_{n=0}^{4L_a-1} \varepsilon_{k,n}(t,\{ \Theta_i (t) \}) d_{k,n}^\dagger d_{k,n}, \\
  &\left( 
    \begin{array}{c}
      d_{k,0} \\
      d_{k,1} \\
      \vdots \\
      d_{4L_a-1}
    \end{array}
  \right) = U_k(t, \{ \Theta_i (t) \})^\dagger \Phi_k.
\end{align}
Then, we can introduce the single particle eigenstates $\Ket{\phi_{k,n}(t)}$ 
with the eigenenergy $\varepsilon_{k,n} (t, \{ \Theta_i (t) \})$ as
\begin{align}
  &H_k^{\text{MF}} \Ket{\phi_{k,n}(t)} = \varepsilon_{k,n} \Ket{\phi_{k,n}(t)}.
\end{align}
In our calculations, we assume that $h_L(-\infty)=0$ and that the system is in the ground state at $t=-\infty$.
Thus, 
the ground state is determined self-consistently as $|\Psi(-\infty)\rangle$, 
which is the many-body state composed of the one-body states $\Ket{\phi_{k,n}(-\infty)}$ with $n$ satisfying $\varepsilon_{k,n}<0$.
This set of $n$ is referred to as ${\cal N}_{\rm in}$.
We regard $|\Psi(-\infty)\rangle$ as an initial state.

In the mean-field theory, the time evolution of 
the one-body wave function $\Ket{\psi_{k,n}(t)}$ is described by the Schr\"{o}dinger equation,
\begin{align}
i \frac{d}{d t} \Ket{\psi_{k,n}(t)} = H^\text{MF}_k(t, \{ \Theta_i (t) \} )\Ket{\psi_{k,n}(t)}, \label{eq:ap1}
\end{align}
where $H^\text{MF}_k$ depends on $\Theta_i$, which is calculated from 
$|\Psi(t)\rangle=\bigotimes_{k,n\in{\cal N}_{\rm in} } \Ket{\psi_{k,n}(t)}$.
We compute the time evolution of
$\Ket{\psi_{k,n}(t)}$ using the extended Euler method under the initial condition, $\Ket{\psi_{k,n}(-\infty)}=\Ket{\phi_{k,n}(-\infty)}$~
\cite{extend_euler-1, extend_euler-2, extend_euler-3, extend_euler-4, extend_euler-5, extend_euler-6}.
First, we calculate $\Ket{\widetilde{\phi}_{k,n}(t')}$ as
\begin{align}
  H_k^{\text{MF}}(t', \{ \Theta_i(t) \} ) \Ket{\widetilde{\phi}_{k,n}(t')} = \widetilde{\varepsilon}_{k,n}(t') \Ket{\widetilde{\phi}_{k,n}(t')},
\end{align}
where $t' = t + \Delta t$.
Then, we calculate 
$\Ket{\widetilde{\psi}_{k,n}(t')}$ as
\begin{align}
  \Ket{\widetilde{\psi}_{k,n}(t')} = \sum_m &e^{-i\Delta t \widetilde{\varepsilon}_{k,m}(t')} \notag \\
  &\times \Braket{\widetilde{\phi}_{k,m}(t') |\psi_{k,n}(t)} \Ket{\widetilde{\phi}_{k,m}(t')}. \label{eq:ap2}
\end{align}

Next, we obtain mean fields $\widetilde{\Theta}_i(t')$ using 
$|\widetilde{\Psi}(t')\rangle=\bigotimes_{k,n\in{\cal N}_{\rm in} } \Ket{\widetilde{\psi}_{k,n}(t')}$,
and $\Ket{\phi_{k,n}(t')}$ as
\begin{align}
  \frac{1}{2} &\left[ H_k^{\text{MF}}(t, \{ \Theta_i(t) \}) + H_k^{\text{MF}}(t', \{ \widetilde{\Theta}_i(t') \} ) \right] \Ket{\phi_{k,n}(t')} \notag \\
  &= \varepsilon_{k,n}(t') \Ket{\phi_{k,n}(t')}.
\end{align}
Finally, we calculate 
$\Ket{\psi_{k,n}(t')}$ in the same manner as Eq.~\eqref{eq:ap2} using $\Ket{\phi_{k,n}(t')}$ and $\varepsilon_{k,n}(t')$,
and mean fields $\Theta_i (t')$ from 
$|\Psi(t')\rangle=\bigotimes_{k,n\in{\cal N}_{\rm in} } \Ket{\psi_{k,n}(t')}$.

By the above procedure, we can obtain mean fields and wave function at $t = t'$.
In order to have the converged solution, we need to take $\Delta t$ small enough.

\bibliography{main}

\begin{thebibliography}{51}%
\makeatletter
\providecommand \@ifxundefined [1]{%
 \@ifx{#1\undefined}
}%
\providecommand \@ifnum [1]{%
 \ifnum #1\expandafter \@firstoftwo
 \else \expandafter \@secondoftwo
 \fi
}%
\providecommand \@ifx [1]{%
 \ifx #1\expandafter \@firstoftwo
 \else \expandafter \@secondoftwo
 \fi
}%
\providecommand \natexlab [1]{#1}%
\providecommand \enquote  [1]{``#1''}%
\providecommand \bibnamefont  [1]{#1}%
\providecommand \bibfnamefont [1]{#1}%
\providecommand \citenamefont [1]{#1}%
\providecommand \href@noop [0]{\@secondoftwo}%
\providecommand \href [0]{\begingroup \@sanitize@url \@href}%
\providecommand \@href[1]{\@@startlink{#1}\@@href}%
\providecommand \@@href[1]{\endgroup#1\@@endlink}%
\providecommand \@sanitize@url [0]{\catcode `\\12\catcode `\$12\catcode
  `\&12\catcode `\#12\catcode `\^12\catcode `\_12\catcode `\%12\relax}%
\providecommand \@@startlink[1]{}%
\providecommand \@@endlink[0]{}%
\providecommand \url  [0]{\begingroup\@sanitize@url \@url }%
\providecommand \@url [1]{\endgroup\@href {#1}{\urlprefix }}%
\providecommand \urlprefix  [0]{URL }%
\providecommand \Eprint [0]{\href }%
\providecommand \doibase [0]{http://dx.doi.org/}%
\providecommand \selectlanguage [0]{\@gobble}%
\providecommand \bibinfo  [0]{\@secondoftwo}%
\providecommand \bibfield  [0]{\@secondoftwo}%
\providecommand \translation [1]{[#1]}%
\providecommand \BibitemOpen [0]{}%
\providecommand \bibitemStop [0]{}%
\providecommand \bibitemNoStop [0]{.\EOS\space}%
\providecommand \EOS [0]{\spacefactor3000\relax}%
\providecommand \BibitemShut  [1]{\csname bibitem#1\endcsname}%
\let\auto@bib@innerbib\@empty
\bibitem [{\citenamefont {Berger}(1996)}]{PhysRevB.54.9353}%
  \BibitemOpen
  \bibfield  {author} {\bibinfo {author} {\bibfnamefont {L.}~\bibnamefont
  {Berger}},\ }\href {\doibase 10.1103/PhysRevB.54.9353} {\bibfield  {journal}
  {\bibinfo  {journal} {Phys. Rev. B}\ }\textbf {\bibinfo {volume} {54}},\
  \bibinfo {pages} {9353} (\bibinfo {year} {1996})}\BibitemShut {NoStop}%
\bibitem [{\citenamefont {Bhat}\ and\ \citenamefont
  {Sipe}(2000)}]{PhysRevLett.85.5432}%
  \BibitemOpen
  \bibfield  {author} {\bibinfo {author} {\bibfnamefont {R.~D.~R.}\
  \bibnamefont {Bhat}}\ and\ \bibinfo {author} {\bibfnamefont {J.~E.}\
  \bibnamefont {Sipe}},\ }\href {\doibase 10.1103/PhysRevLett.85.5432}
  {\bibfield  {journal} {\bibinfo  {journal} {Phys. Rev. Lett.}\ }\textbf
  {\bibinfo {volume} {85}},\ \bibinfo {pages} {5432} (\bibinfo {year}
  {2000})}\BibitemShut {NoStop}%
\bibitem [{\citenamefont {\ifmmode \check{Z}\else
  \v{Z}\fi{}uti\ifmmode~\acute{c}\else \'{c}\fi{}}\ \emph
  {et~al.}(2004)\citenamefont {\ifmmode \check{Z}\else
  \v{Z}\fi{}uti\ifmmode~\acute{c}\else \'{c}\fi{}}, \citenamefont {Fabian},\
  and\ \citenamefont {Das~Sarma}}]{Spintronics}%
  \BibitemOpen
  \bibfield  {author} {\bibinfo {author} {\bibfnamefont {I.}~\bibnamefont
  {\ifmmode \check{Z}\else \v{Z}\fi{}uti\ifmmode~\acute{c}\else \'{c}\fi{}}},
  \bibinfo {author} {\bibfnamefont {J.}~\bibnamefont {Fabian}}, \ and\ \bibinfo
  {author} {\bibfnamefont {S.}~\bibnamefont {Das~Sarma}},\ }\href {\doibase
  10.1103/RevModPhys.76.323} {\bibfield  {journal} {\bibinfo  {journal} {Rev.
  Mod. Phys.}\ }\textbf {\bibinfo {volume} {76}},\ \bibinfo {pages} {323}
  (\bibinfo {year} {2004})}\BibitemShut {NoStop}%
\bibitem [{\citenamefont {Tsoi}\ \emph {et~al.}(2000)\citenamefont {Tsoi},
  \citenamefont {Jansen}, \citenamefont {Bass}, \citenamefont {Chiang},
  \citenamefont {Tsoi},\ and\ \citenamefont {Wyder}}]{Tsoi2000}%
  \BibitemOpen
  \bibfield  {author} {\bibinfo {author} {\bibfnamefont {M.}~\bibnamefont
  {Tsoi}}, \bibinfo {author} {\bibfnamefont {A.~G.~M.}\ \bibnamefont {Jansen}},
  \bibinfo {author} {\bibfnamefont {J.}~\bibnamefont {Bass}}, \bibinfo {author}
  {\bibfnamefont {W.-C.}\ \bibnamefont {Chiang}}, \bibinfo {author}
  {\bibfnamefont {V.}~\bibnamefont {Tsoi}}, \ and\ \bibinfo {author}
  {\bibfnamefont {P.}~\bibnamefont {Wyder}},\ }\href {\doibase
  10.1038/35017512} {\bibfield  {journal} {\bibinfo  {journal} {Nature}\
  }\textbf {\bibinfo {volume} {406}},\ \bibinfo {pages} {46} (\bibinfo {year}
  {2000})}\BibitemShut {NoStop}%
\bibitem [{\citenamefont {Slonczewski}(1996)}]{SpinCurrent-1}%
  \BibitemOpen
  \bibfield  {author} {\bibinfo {author} {\bibfnamefont {J.}~\bibnamefont
  {Slonczewski}},\ }\href {\doibase
  https://doi.org/10.1016/0304-8853(96)00062-5} {\bibfield  {journal} {\bibinfo
   {journal} {J. Magn. Magn. Mater.}\ }\textbf {\bibinfo {volume} {159}},\
  \bibinfo {pages} {L1} (\bibinfo {year} {1996})}\BibitemShut {NoStop}%
\bibitem [{\citenamefont {K\"onig}\ \emph {et~al.}(2001)\citenamefont
  {K\"onig}, \citenamefont {B\o{}nsager},\ and\ \citenamefont
  {MacDonald}}]{PhysRevLett.87.187202}%
  \BibitemOpen
  \bibfield  {author} {\bibinfo {author} {\bibfnamefont {J.}~\bibnamefont
  {K\"onig}}, \bibinfo {author} {\bibfnamefont {M.~C.}\ \bibnamefont
  {B\o{}nsager}}, \ and\ \bibinfo {author} {\bibfnamefont {A.~H.}\ \bibnamefont
  {MacDonald}},\ }\href {\doibase 10.1103/PhysRevLett.87.187202} {\bibfield
  {journal} {\bibinfo  {journal} {Phys. Rev. Lett.}\ }\textbf {\bibinfo
  {volume} {87}},\ \bibinfo {pages} {187202} (\bibinfo {year}
  {2001})}\BibitemShut {NoStop}%
\bibitem [{\citenamefont {Slonczewski}(1989)}]{PhysRevB.39.6995}%
  \BibitemOpen
  \bibfield  {author} {\bibinfo {author} {\bibfnamefont {J.~C.}\ \bibnamefont
  {Slonczewski}},\ }\href {\doibase 10.1103/PhysRevB.39.6995} {\bibfield
  {journal} {\bibinfo  {journal} {Phys. Rev. B}\ }\textbf {\bibinfo {volume}
  {39}},\ \bibinfo {pages} {6995} (\bibinfo {year} {1989})}\BibitemShut
  {NoStop}%
\bibitem [{\citenamefont {Ogawa}\ \emph {et~al.}(2016)\citenamefont {Ogawa},
  \citenamefont {Yoshimi}, \citenamefont {Yasuda}, \citenamefont {Tsukazaki},
  \citenamefont {Kawasaki},\ and\ \citenamefont {Tokura}}]{Ogawa2016}%
  \BibitemOpen
  \bibfield  {author} {\bibinfo {author} {\bibfnamefont {N.}~\bibnamefont
  {Ogawa}}, \bibinfo {author} {\bibfnamefont {R.}~\bibnamefont {Yoshimi}},
  \bibinfo {author} {\bibfnamefont {K.}~\bibnamefont {Yasuda}}, \bibinfo
  {author} {\bibfnamefont {A.}~\bibnamefont {Tsukazaki}}, \bibinfo {author}
  {\bibfnamefont {M.}~\bibnamefont {Kawasaki}}, \ and\ \bibinfo {author}
  {\bibfnamefont {Y.}~\bibnamefont {Tokura}},\ }\href {\doibase
  10.1038/ncomms12246} {\bibfield  {journal} {\bibinfo  {journal} {Nat.
  Commun.}\ }\textbf {\bibinfo {volume} {7}},\ \bibinfo {pages} {12246}
  (\bibinfo {year} {2016})}\BibitemShut {NoStop}%
\bibitem [{\citenamefont {Tsui}\ \emph {et~al.}(1971)\citenamefont {Tsui},
  \citenamefont {Dietz},\ and\ \citenamefont {Walker}}]{PhysRevLett.27.1729}%
  \BibitemOpen
  \bibfield  {author} {\bibinfo {author} {\bibfnamefont {D.~C.}\ \bibnamefont
  {Tsui}}, \bibinfo {author} {\bibfnamefont {R.~E.}\ \bibnamefont {Dietz}}, \
  and\ \bibinfo {author} {\bibfnamefont {L.~R.}\ \bibnamefont {Walker}},\
  }\href {\doibase 10.1103/PhysRevLett.27.1729} {\bibfield  {journal} {\bibinfo
   {journal} {Phys. Rev. Lett.}\ }\textbf {\bibinfo {volume} {27}},\ \bibinfo
  {pages} {1729} (\bibinfo {year} {1971})}\BibitemShut {NoStop}%
\bibitem [{\citenamefont {Moodera}\ \emph {et~al.}(1995)\citenamefont
  {Moodera}, \citenamefont {Kinder}, \citenamefont {Wong},\ and\ \citenamefont
  {Meservey}}]{PhysRevLett.74.3273}%
  \BibitemOpen
  \bibfield  {author} {\bibinfo {author} {\bibfnamefont {J.~S.}\ \bibnamefont
  {Moodera}}, \bibinfo {author} {\bibfnamefont {L.~R.}\ \bibnamefont {Kinder}},
  \bibinfo {author} {\bibfnamefont {T.~M.}\ \bibnamefont {Wong}}, \ and\
  \bibinfo {author} {\bibfnamefont {R.}~\bibnamefont {Meservey}},\ }\href
  {\doibase 10.1103/PhysRevLett.74.3273} {\bibfield  {journal} {\bibinfo
  {journal} {Phys. Rev. Lett.}\ }\textbf {\bibinfo {volume} {74}},\ \bibinfo
  {pages} {3273} (\bibinfo {year} {1995})}\BibitemShut {NoStop}%
\bibitem [{\citenamefont {Kajiwara}\ \emph {et~al.}(2010)\citenamefont
  {Kajiwara}, \citenamefont {Harii}, \citenamefont {Takahashi}, \citenamefont
  {Ohe}, \citenamefont {Uchida}, \citenamefont {Mizuguchi}, \citenamefont
  {Umezawa}, \citenamefont {Kawai}, \citenamefont {Ando}, \citenamefont
  {Takanashi}, \citenamefont {Maekawa},\ and\ \citenamefont
  {Saitoh}}]{SpinCurrent-2}%
  \BibitemOpen
  \bibfield  {author} {\bibinfo {author} {\bibfnamefont {Y.}~\bibnamefont
  {Kajiwara}}, \bibinfo {author} {\bibfnamefont {K.}~\bibnamefont {Harii}},
  \bibinfo {author} {\bibfnamefont {S.}~\bibnamefont {Takahashi}}, \bibinfo
  {author} {\bibfnamefont {J.}~\bibnamefont {Ohe}}, \bibinfo {author}
  {\bibfnamefont {K.}~\bibnamefont {Uchida}}, \bibinfo {author} {\bibfnamefont
  {M.}~\bibnamefont {Mizuguchi}}, \bibinfo {author} {\bibfnamefont
  {H.}~\bibnamefont {Umezawa}}, \bibinfo {author} {\bibfnamefont
  {H.}~\bibnamefont {Kawai}}, \bibinfo {author} {\bibfnamefont
  {K.}~\bibnamefont {Ando}}, \bibinfo {author} {\bibfnamefont {K.}~\bibnamefont
  {Takanashi}}, \bibinfo {author} {\bibfnamefont {S.}~\bibnamefont {Maekawa}},
  \ and\ \bibinfo {author} {\bibfnamefont {E.}~\bibnamefont {Saitoh}},\ }\href
  {\doibase 10.1038/nature08876} {\bibfield  {journal} {\bibinfo  {journal}
  {Nature}\ }\textbf {\bibinfo {volume} {464}},\ \bibinfo {pages} {262}
  (\bibinfo {year} {2010})}\BibitemShut {NoStop}%
\bibitem [{\citenamefont {Cornelissen}\ \emph {et~al.}(2015)\citenamefont
  {Cornelissen}, \citenamefont {Liu}, \citenamefont {Duine}, \citenamefont
  {Youssef},\ and\ \citenamefont {van Wees}}]{SpinCurrent-3}%
  \BibitemOpen
  \bibfield  {author} {\bibinfo {author} {\bibfnamefont {L.~J.}\ \bibnamefont
  {Cornelissen}}, \bibinfo {author} {\bibfnamefont {J.}~\bibnamefont {Liu}},
  \bibinfo {author} {\bibfnamefont {R.~A.}\ \bibnamefont {Duine}}, \bibinfo
  {author} {\bibfnamefont {J.~B.}\ \bibnamefont {Youssef}}, \ and\ \bibinfo
  {author} {\bibfnamefont {B.~J.}\ \bibnamefont {van Wees}},\ }\href {\doibase
  10.1038/nphys3465} {\bibfield  {journal} {\bibinfo  {journal} {Nat. Phys.}\
  }\textbf {\bibinfo {volume} {11}},\ \bibinfo {pages} {1022} (\bibinfo {year}
  {2015})}\BibitemShut {NoStop}%
\bibitem [{\citenamefont {Anderson}(1973)}]{ANDERSON1973153}%
  \BibitemOpen
  \bibfield  {author} {\bibinfo {author} {\bibfnamefont {P.}~\bibnamefont
  {Anderson}},\ }\href {\doibase https://doi.org/10.1016/0025-5408(73)90167-0}
  {\bibfield  {journal} {\bibinfo  {journal} {Mater. Res. Bull.}\ }\textbf
  {\bibinfo {volume} {8}},\ \bibinfo {pages} {153} (\bibinfo {year}
  {1973})}\BibitemShut {NoStop}%
\bibitem [{\citenamefont {Savary}\ and\ \citenamefont
  {Balents}(2016{\natexlab{a}})}]{Savary_2016}%
  \BibitemOpen
  \bibfield  {author} {\bibinfo {author} {\bibfnamefont {L.}~\bibnamefont
  {Savary}}\ and\ \bibinfo {author} {\bibfnamefont {L.}~\bibnamefont
  {Balents}},\ }\href {\doibase 10.1088/0034-4885/80/1/016502} {\bibfield
  {journal} {\bibinfo  {journal} {Rep. Prog. Phys.}\ }\textbf {\bibinfo
  {volume} {80}},\ \bibinfo {pages} {016502} (\bibinfo {year}
  {2016}{\natexlab{a}})}\BibitemShut {NoStop}%
\bibitem [{\citenamefont {Balents}(2010)}]{qsl1}%
  \BibitemOpen
  \bibfield  {author} {\bibinfo {author} {\bibfnamefont {L.}~\bibnamefont
  {Balents}},\ }\href {\doibase 10.1038/nature08917} {\bibfield  {journal}
  {\bibinfo  {journal} {Nat. Phys.}\ }\textbf {\bibinfo {volume} {464}},\
  \bibinfo {pages} {199} (\bibinfo {year} {2010})}\BibitemShut {NoStop}%
\bibitem [{\citenamefont {Savary}\ and\ \citenamefont
  {Balents}(2016{\natexlab{b}})}]{qsl2}%
  \BibitemOpen
  \bibfield  {author} {\bibinfo {author} {\bibfnamefont {L.}~\bibnamefont
  {Savary}}\ and\ \bibinfo {author} {\bibfnamefont {L.}~\bibnamefont
  {Balents}},\ }\href {\doibase 10.1088/0034-4885/80/1/016502} {\bibfield
  {journal} {\bibinfo  {journal} {Rep. Prog. Phys,}\ }\textbf {\bibinfo
  {volume} {80}},\ \bibinfo {pages} {016502} (\bibinfo {year}
  {2016}{\natexlab{b}})}\BibitemShut {NoStop}%
\bibitem [{\citenamefont {Chen}\ \emph {et~al.}(2013)\citenamefont {Chen},
  \citenamefont {Sun}, \citenamefont {Wang},\ and\ \citenamefont
  {Xie}}]{PhysRevB.88.041405}%
  \BibitemOpen
  \bibfield  {author} {\bibinfo {author} {\bibfnamefont {C.-Z.}\ \bibnamefont
  {Chen}}, \bibinfo {author} {\bibfnamefont {Q.-f.}\ \bibnamefont {Sun}},
  \bibinfo {author} {\bibfnamefont {F.}~\bibnamefont {Wang}}, \ and\ \bibinfo
  {author} {\bibfnamefont {X.~C.}\ \bibnamefont {Xie}},\ }\href {\doibase
  10.1103/PhysRevB.88.041405} {\bibfield  {journal} {\bibinfo  {journal} {Phys.
  Rev. B}\ }\textbf {\bibinfo {volume} {88}},\ \bibinfo {pages} {041405}
  (\bibinfo {year} {2013})}\BibitemShut {NoStop}%
\bibitem [{\citenamefont {Read}\ and\ \citenamefont
  {Chakraborty}(1989)}]{PhysRevB.40.7133}%
  \BibitemOpen
  \bibfield  {author} {\bibinfo {author} {\bibfnamefont {N.}~\bibnamefont
  {Read}}\ and\ \bibinfo {author} {\bibfnamefont {B.}~\bibnamefont
  {Chakraborty}},\ }\href {\doibase 10.1103/PhysRevB.40.7133} {\bibfield
  {journal} {\bibinfo  {journal} {Phys. Rev. B}\ }\textbf {\bibinfo {volume}
  {40}},\ \bibinfo {pages} {7133} (\bibinfo {year} {1989})}\BibitemShut
  {NoStop}%
\bibitem [{\citenamefont {Wen}(1991)}]{PhysRevB.44.2664}%
  \BibitemOpen
  \bibfield  {author} {\bibinfo {author} {\bibfnamefont {X.~G.}\ \bibnamefont
  {Wen}},\ }\href {\doibase 10.1103/PhysRevB.44.2664} {\bibfield  {journal}
  {\bibinfo  {journal} {Phys. Rev. B}\ }\textbf {\bibinfo {volume} {44}},\
  \bibinfo {pages} {2664} (\bibinfo {year} {1991})}\BibitemShut {NoStop}%
\bibitem [{\citenamefont {Zhou}\ \emph {et~al.}(2017)\citenamefont {Zhou},
  \citenamefont {Kanoda},\ and\ \citenamefont {Ng}}]{RevModPhys.89.025003}%
  \BibitemOpen
  \bibfield  {author} {\bibinfo {author} {\bibfnamefont {Y.}~\bibnamefont
  {Zhou}}, \bibinfo {author} {\bibfnamefont {K.}~\bibnamefont {Kanoda}}, \ and\
  \bibinfo {author} {\bibfnamefont {T.-K.}\ \bibnamefont {Ng}},\ }\href
  {\doibase 10.1103/RevModPhys.89.025003} {\bibfield  {journal} {\bibinfo
  {journal} {Rev. Mod. Phys.}\ }\textbf {\bibinfo {volume} {89}},\ \bibinfo
  {pages} {025003} (\bibinfo {year} {2017})}\BibitemShut {NoStop}%
\bibitem [{\citenamefont {Hirobe}\ \emph {et~al.}(2017)\citenamefont {Hirobe},
  \citenamefont {Sato}, \citenamefont {Kawamata}, \citenamefont {Shiomi},
  \citenamefont {i.~Uchida}, \citenamefont {Iguchi}, \citenamefont {Koike},
  \citenamefont {Maekawa}, ,\ and\ \citenamefont {Saitoh}}]{SpinonSpinCurrent}%
  \BibitemOpen
  \bibfield  {author} {\bibinfo {author} {\bibfnamefont {D.}~\bibnamefont
  {Hirobe}}, \bibinfo {author} {\bibfnamefont {M.}~\bibnamefont {Sato}},
  \bibinfo {author} {\bibfnamefont {T.}~\bibnamefont {Kawamata}}, \bibinfo
  {author} {\bibfnamefont {Y.}~\bibnamefont {Shiomi}}, \bibinfo {author}
  {\bibfnamefont {K.}~\bibnamefont {i.~Uchida}}, \bibinfo {author}
  {\bibfnamefont {R.}~\bibnamefont {Iguchi}}, \bibinfo {author} {\bibfnamefont
  {Y.}~\bibnamefont {Koike}}, \bibinfo {author} {\bibfnamefont
  {S.}~\bibnamefont {Maekawa}}, , \ and\ \bibinfo {author} {\bibfnamefont
  {E.}~\bibnamefont {Saitoh}},\ }\href {\doibase 10.1038/nphys3895} {\bibfield
  {journal} {\bibinfo  {journal} {Nat. Phys.}\ }\textbf {\bibinfo {volume}
  {13}},\ \bibinfo {pages} {30} (\bibinfo {year} {2017})}\BibitemShut {NoStop}%
\bibitem [{\citenamefont {Kitaev}(2006)}]{Kitaev_model}%
  \BibitemOpen
  \bibfield  {author} {\bibinfo {author} {\bibfnamefont {A.}~\bibnamefont
  {Kitaev}},\ }\href {\doibase https://doi.org/10.1016/j.aop.2005.10.005}
  {\bibfield  {journal} {\bibinfo  {journal} {Ann. Phys.}\ }\textbf {\bibinfo
  {volume} {321}},\ \bibinfo {pages} {2 } (\bibinfo {year} {2006})}\BibitemShut
  {NoStop}%
\bibitem [{\citenamefont {Chen}\ and\ \citenamefont
  {Hu}(2007)}]{PhysRevB.76.193101}%
  \BibitemOpen
  \bibfield  {author} {\bibinfo {author} {\bibfnamefont {H.-D.}\ \bibnamefont
  {Chen}}\ and\ \bibinfo {author} {\bibfnamefont {J.}~\bibnamefont {Hu}},\
  }\href {\doibase 10.1103/PhysRevB.76.193101} {\bibfield  {journal} {\bibinfo
  {journal} {Phys. Rev. B}\ }\textbf {\bibinfo {volume} {76}},\ \bibinfo
  {pages} {193101} (\bibinfo {year} {2007})}\BibitemShut {NoStop}%
\bibitem [{\citenamefont {Feng}\ \emph {et~al.}(2007)\citenamefont {Feng},
  \citenamefont {Zhang},\ and\ \citenamefont {Xiang}}]{PhysRevLett.98.087204}%
  \BibitemOpen
  \bibfield  {author} {\bibinfo {author} {\bibfnamefont {X.-Y.}\ \bibnamefont
  {Feng}}, \bibinfo {author} {\bibfnamefont {G.-M.}\ \bibnamefont {Zhang}}, \
  and\ \bibinfo {author} {\bibfnamefont {T.}~\bibnamefont {Xiang}},\ }\href
  {\doibase 10.1103/PhysRevLett.98.087204} {\bibfield  {journal} {\bibinfo
  {journal} {Phys. Rev. Lett.}\ }\textbf {\bibinfo {volume} {98}},\ \bibinfo
  {pages} {087204} (\bibinfo {year} {2007})}\BibitemShut {NoStop}%
\bibitem [{\citenamefont {Chen}\ and\ \citenamefont
  {Nussinov}(2008)}]{Kitaev_review}%
  \BibitemOpen
  \bibfield  {author} {\bibinfo {author} {\bibfnamefont {H.-D.}\ \bibnamefont
  {Chen}}\ and\ \bibinfo {author} {\bibfnamefont {Z.}~\bibnamefont
  {Nussinov}},\ }\href {\doibase 10.1088/1751-8113/41/7/075001} {\bibfield
  {journal} {\bibinfo  {journal} {J. Phys. A: Math. Theor.}\ }\textbf {\bibinfo
  {volume} {41}},\ \bibinfo {pages} {075001} (\bibinfo {year}
  {2008})}\BibitemShut {NoStop}%
\bibitem [{\citenamefont {Motome}\ and\ \citenamefont
  {Nasu}(2020)}]{Motome_Nasu_Review}%
  \BibitemOpen
  \bibfield  {author} {\bibinfo {author} {\bibfnamefont {Y.}~\bibnamefont
  {Motome}}\ and\ \bibinfo {author} {\bibfnamefont {J.}~\bibnamefont {Nasu}},\
  }\href {\doibase 10.7566/JPSJ.89.012002} {\bibfield  {journal} {\bibinfo
  {journal} {J. Phys. Soc. Jpn}\ }\textbf {\bibinfo {volume} {89}},\ \bibinfo
  {pages} {012002} (\bibinfo {year} {2020})}\BibitemShut {NoStop}%
\bibitem [{\citenamefont {Minakawa}\ \emph {et~al.}(2020)\citenamefont
  {Minakawa}, \citenamefont {Murakami}, \citenamefont {Koga},\ and\
  \citenamefont {Nasu}}]{spin_trans}%
  \BibitemOpen
  \bibfield  {author} {\bibinfo {author} {\bibfnamefont {T.}~\bibnamefont
  {Minakawa}}, \bibinfo {author} {\bibfnamefont {Y.}~\bibnamefont {Murakami}},
  \bibinfo {author} {\bibfnamefont {A.}~\bibnamefont {Koga}}, \ and\ \bibinfo
  {author} {\bibfnamefont {J.}~\bibnamefont {Nasu}},\ }\href {\doibase
  10.1103/PhysRevLett.125.047204} {\bibfield  {journal} {\bibinfo  {journal}
  {Phys. Rev. Lett.}\ }\textbf {\bibinfo {volume} {125}},\ \bibinfo {pages}
  {047204} (\bibinfo {year} {2020})}\BibitemShut {NoStop}%
\bibitem [{\citenamefont {Koga}\ \emph {et~al.}(2020)\citenamefont {Koga},
  \citenamefont {Minakawa}, \citenamefont {Murakami},\ and\ \citenamefont
  {Nasu}}]{spin_trans_2}%
  \BibitemOpen
  \bibfield  {author} {\bibinfo {author} {\bibfnamefont {A.}~\bibnamefont
  {Koga}}, \bibinfo {author} {\bibfnamefont {T.}~\bibnamefont {Minakawa}},
  \bibinfo {author} {\bibfnamefont {Y.}~\bibnamefont {Murakami}}, \ and\
  \bibinfo {author} {\bibfnamefont {J.}~\bibnamefont {Nasu}},\ }\href {\doibase
  10.7566/JPSJ.89.033701} {\bibfield  {journal} {\bibinfo  {journal} {J. Phys.
  Soc. Jpn}\ }\textbf {\bibinfo {volume} {89}},\ \bibinfo {pages} {033701}
  (\bibinfo {year} {2020})}\BibitemShut {NoStop}%
\bibitem [{\citenamefont {Nasu}\ \emph {et~al.}(2018)\citenamefont {Nasu},
  \citenamefont {Kato}, \citenamefont {Kamiya},\ and\ \citenamefont
  {Motome}}]{Nasu_2018}%
  \BibitemOpen
  \bibfield  {author} {\bibinfo {author} {\bibfnamefont {J.}~\bibnamefont
  {Nasu}}, \bibinfo {author} {\bibfnamefont {Y.}~\bibnamefont {Kato}}, \bibinfo
  {author} {\bibfnamefont {Y.}~\bibnamefont {Kamiya}}, \ and\ \bibinfo {author}
  {\bibfnamefont {Y.}~\bibnamefont {Motome}},\ }\href {\doibase
  10.1103/PhysRevB.98.060416} {\bibfield  {journal} {\bibinfo  {journal} {Phys.
  Rev. B}\ }\textbf {\bibinfo {volume} {98}},\ \bibinfo {pages} {060416(R)}
  (\bibinfo {year} {2018})}\BibitemShut {NoStop}%
\bibitem [{\citenamefont {Nasu}\ and\ \citenamefont
  {Motome}(2019)}]{MajoranaMF-1}%
  \BibitemOpen
  \bibfield  {author} {\bibinfo {author} {\bibfnamefont {J.}~\bibnamefont
  {Nasu}}\ and\ \bibinfo {author} {\bibfnamefont {Y.}~\bibnamefont {Motome}},\
  }\href {\doibase 10.1103/PhysRevResearch.1.033007} {\bibfield  {journal}
  {\bibinfo  {journal} {Phys. Rev. Research}\ }\textbf {\bibinfo {volume}
  {1}},\ \bibinfo {pages} {033007} (\bibinfo {year} {2019})}\BibitemShut
  {NoStop}%
\bibitem [{\citenamefont {Nasu}\ \emph {et~al.}(2014)\citenamefont {Nasu},
  \citenamefont {Udagawa},\ and\ \citenamefont
  {Motome}}]{PhysRevLett.113.197205}%
  \BibitemOpen
  \bibfield  {author} {\bibinfo {author} {\bibfnamefont {J.}~\bibnamefont
  {Nasu}}, \bibinfo {author} {\bibfnamefont {M.}~\bibnamefont {Udagawa}}, \
  and\ \bibinfo {author} {\bibfnamefont {Y.}~\bibnamefont {Motome}},\ }\href
  {\doibase 10.1103/PhysRevLett.113.197205} {\bibfield  {journal} {\bibinfo
  {journal} {Phys. Rev. Lett.}\ }\textbf {\bibinfo {volume} {113}},\ \bibinfo
  {pages} {197205} (\bibinfo {year} {2014})}\BibitemShut {NoStop}%
\bibitem [{\citenamefont {Majorana}(1937)}]{Majorana2008}%
  \BibitemOpen
  \bibfield  {author} {\bibinfo {author} {\bibfnamefont {E.}~\bibnamefont
  {Majorana}},\ }\href {\doibase 10.1007/BF02961314} {\bibfield  {journal}
  {\bibinfo  {journal} {Il Nuovo Cimento}\ }\textbf {\bibinfo {volume} {14}},\
  \bibinfo {pages} {171} (\bibinfo {year} {1937})}\BibitemShut {NoStop}%
\bibitem [{\citenamefont {Wilczek}(2009)}]{Wilczek2009}%
  \BibitemOpen
  \bibfield  {author} {\bibinfo {author} {\bibfnamefont {F.}~\bibnamefont
  {Wilczek}},\ }\href {\doibase 10.1038/nphys1380} {\bibfield  {journal}
  {\bibinfo  {journal} {Nat. Phys.}\ }\textbf {\bibinfo {volume} {5}},\
  \bibinfo {pages} {614} (\bibinfo {year} {2009})}\BibitemShut {NoStop}%
\bibitem [{\citenamefont {Terai}\ and\ \citenamefont
  {Ono}(1993)}]{extend_euler-1}%
  \BibitemOpen
  \bibfield  {author} {\bibinfo {author} {\bibfnamefont {A.}~\bibnamefont
  {Terai}}\ and\ \bibinfo {author} {\bibfnamefont {Y.}~\bibnamefont {Ono}},\
  }\href {\doibase 10.1143/PTPS.113.177} {\bibfield  {journal} {\bibinfo
  {journal} {Prog. Theor. Phys. Suppl.}\ }\textbf {\bibinfo {volume} {113}},\
  \bibinfo {pages} {177} (\bibinfo {year} {1993})}\BibitemShut {NoStop}%
\bibitem [{\citenamefont {Hirano}\ and\ \citenamefont
  {Ono}(2000)}]{extend_euler-2}%
  \BibitemOpen
  \bibfield  {author} {\bibinfo {author} {\bibfnamefont {Y.}~\bibnamefont
  {Hirano}}\ and\ \bibinfo {author} {\bibfnamefont {Y.}~\bibnamefont {Ono}},\
  }\href {\doibase 10.1143/JPSJ.69.2131} {\bibfield  {journal} {\bibinfo
  {journal} {J. Phys. Soc. Jpn.}\ }\textbf {\bibinfo {volume} {69}},\ \bibinfo
  {pages} {2131} (\bibinfo {year} {2000})}\BibitemShut {NoStop}%
\bibitem [{\citenamefont {Tanaka}\ and\ \citenamefont
  {Yonemitsu}(2010)}]{extend_euler-3}%
  \BibitemOpen
  \bibfield  {author} {\bibinfo {author} {\bibfnamefont {Y.}~\bibnamefont
  {Tanaka}}\ and\ \bibinfo {author} {\bibfnamefont {K.}~\bibnamefont
  {Yonemitsu}},\ }\href {\doibase 10.1143/JPSJ.79.024712} {\bibfield  {journal}
  {\bibinfo  {journal} {J. Phys. Soc. Jpn}\ }\textbf {\bibinfo {volume} {79}},\
  \bibinfo {pages} {024712} (\bibinfo {year} {2010})}\BibitemShut {NoStop}%
\bibitem [{\citenamefont {Ohara}\ and\ \citenamefont
  {Yamamoto}(2017)}]{extend_euler-4}%
  \BibitemOpen
  \bibfield  {author} {\bibinfo {author} {\bibfnamefont {J.}~\bibnamefont
  {Ohara}}\ and\ \bibinfo {author} {\bibfnamefont {S.}~\bibnamefont
  {Yamamoto}},\ }\href {\doibase 10.1088/1742-6596/868/1/012013} {\bibfield
  {journal} {\bibinfo  {journal} {J. Phys.:Conf. Ser.}\ }\textbf {\bibinfo
  {volume} {868}},\ \bibinfo {pages} {012013} (\bibinfo {year}
  {2017})}\BibitemShut {NoStop}%
\bibitem [{\citenamefont {Tanaka}\ \emph {et~al.}(2018)\citenamefont {Tanaka},
  \citenamefont {Daira},\ and\ \citenamefont {Yonemitsu}}]{extend_euler-5}%
  \BibitemOpen
  \bibfield  {author} {\bibinfo {author} {\bibfnamefont {Y.}~\bibnamefont
  {Tanaka}}, \bibinfo {author} {\bibfnamefont {M.}~\bibnamefont {Daira}}, \
  and\ \bibinfo {author} {\bibfnamefont {K.}~\bibnamefont {Yonemitsu}},\ }\href
  {\doibase 10.1103/PhysRevB.97.115105} {\bibfield  {journal} {\bibinfo
  {journal} {Phys. Rev. B}\ }\textbf {\bibinfo {volume} {97}},\ \bibinfo
  {pages} {115105} (\bibinfo {year} {2018})}\BibitemShut {NoStop}%
\bibitem [{\citenamefont {Seo}\ \emph {et~al.}(2018)\citenamefont {Seo},
  \citenamefont {Tanaka},\ and\ \citenamefont {Ishihara}}]{extend_euler-6}%
  \BibitemOpen
  \bibfield  {author} {\bibinfo {author} {\bibfnamefont {H.}~\bibnamefont
  {Seo}}, \bibinfo {author} {\bibfnamefont {Y.}~\bibnamefont {Tanaka}}, \ and\
  \bibinfo {author} {\bibfnamefont {S.}~\bibnamefont {Ishihara}},\ }\href
  {\doibase 10.1103/PhysRevB.98.235150} {\bibfield  {journal} {\bibinfo
  {journal} {Phys. Rev. B}\ }\textbf {\bibinfo {volume} {98}},\ \bibinfo
  {pages} {235150} (\bibinfo {year} {2018})}\BibitemShut {NoStop}%
\bibitem [{\citenamefont {Liang}\ \emph {et~al.}(2018)\citenamefont {Liang},
  \citenamefont {Jiang}, \citenamefont {Chen}, \citenamefont {Li},\ and\
  \citenamefont {Wang}}]{PhysRevB.98.054433}%
  \BibitemOpen
  \bibfield  {author} {\bibinfo {author} {\bibfnamefont {S.}~\bibnamefont
  {Liang}}, \bibinfo {author} {\bibfnamefont {M.-H.}\ \bibnamefont {Jiang}},
  \bibinfo {author} {\bibfnamefont {W.}~\bibnamefont {Chen}}, \bibinfo {author}
  {\bibfnamefont {J.-X.}\ \bibnamefont {Li}}, \ and\ \bibinfo {author}
  {\bibfnamefont {Q.-H.}\ \bibnamefont {Wang}},\ }\href {\doibase
  10.1103/PhysRevB.98.054433} {\bibfield  {journal} {\bibinfo  {journal} {Phys.
  Rev. B}\ }\textbf {\bibinfo {volume} {98}},\ \bibinfo {pages} {054433}
  (\bibinfo {year} {2018})}\BibitemShut {NoStop}%
\bibitem [{\citenamefont {Ido}\ and\ \citenamefont
  {Misawa}(2020)}]{PhysRevB.101.045121}%
  \BibitemOpen
  \bibfield  {author} {\bibinfo {author} {\bibfnamefont {K.}~\bibnamefont
  {Ido}}\ and\ \bibinfo {author} {\bibfnamefont {T.}~\bibnamefont {Misawa}},\
  }\href {\doibase 10.1103/PhysRevB.101.045121} {\bibfield  {journal} {\bibinfo
   {journal} {Phys. Rev. B}\ }\textbf {\bibinfo {volume} {101}},\ \bibinfo
  {pages} {045121} (\bibinfo {year} {2020})}\BibitemShut {NoStop}%
\bibitem [{\citenamefont {Baskaran}\ \emph {et~al.}(2007)\citenamefont
  {Baskaran}, \citenamefont {Mandal},\ and\ \citenamefont
  {Shankar}}]{short_interaction}%
  \BibitemOpen
  \bibfield  {author} {\bibinfo {author} {\bibfnamefont {G.}~\bibnamefont
  {Baskaran}}, \bibinfo {author} {\bibfnamefont {S.}~\bibnamefont {Mandal}}, \
  and\ \bibinfo {author} {\bibfnamefont {R.}~\bibnamefont {Shankar}},\ }\href
  {\doibase 10.1103/PhysRevLett.98.247201} {\bibfield  {journal} {\bibinfo
  {journal} {Phys. Rev. Lett.}\ }\textbf {\bibinfo {volume} {98}},\ \bibinfo
  {pages} {247201} (\bibinfo {year} {2007})}\BibitemShut {NoStop}%
\bibitem [{\citenamefont {Nussinov}\ and\ \citenamefont {van~den
  Brink}(2015)}]{RevModPhys.87.1}%
  \BibitemOpen
  \bibfield  {author} {\bibinfo {author} {\bibfnamefont {Z.}~\bibnamefont
  {Nussinov}}\ and\ \bibinfo {author} {\bibfnamefont {J.}~\bibnamefont {van~den
  Brink}},\ }\href {\doibase 10.1103/RevModPhys.87.1} {\bibfield  {journal}
  {\bibinfo  {journal} {Rev. Mod. Phys.}\ }\textbf {\bibinfo {volume} {87}},\
  \bibinfo {pages} {1} (\bibinfo {year} {2015})}\BibitemShut {NoStop}%
\bibitem [{\citenamefont {Trebst}(2017)}]{KitaevMaterial-1}%
  \BibitemOpen
  \bibfield  {author} {\bibinfo {author} {\bibfnamefont {S.}~\bibnamefont
  {Trebst}},\ }\href@noop {} {\enquote {\bibinfo {title} {Kitaev materials},}\
  } (\bibinfo {year} {2017}),\ \Eprint {http://arxiv.org/abs/1701.07056}
  {arXiv:1701.07056 [cond-mat.str-el]} \BibitemShut {NoStop}%
\bibitem [{\citenamefont {Winter}\ \emph {et~al.}(2017)\citenamefont {Winter},
  \citenamefont {Tsirlin}, \citenamefont {Daghofer}, \citenamefont {van~den
  Brink}, \citenamefont {Singh}, \citenamefont {Gegenwart},\ and\ \citenamefont
  {Valent{\'{\i}}}}]{KitaevMaterial-2}%
  \BibitemOpen
  \bibfield  {author} {\bibinfo {author} {\bibfnamefont {S.~M.}\ \bibnamefont
  {Winter}}, \bibinfo {author} {\bibfnamefont {A.~A.}\ \bibnamefont {Tsirlin}},
  \bibinfo {author} {\bibfnamefont {M.}~\bibnamefont {Daghofer}}, \bibinfo
  {author} {\bibfnamefont {J.}~\bibnamefont {van~den Brink}}, \bibinfo {author}
  {\bibfnamefont {Y.}~\bibnamefont {Singh}}, \bibinfo {author} {\bibfnamefont
  {P.}~\bibnamefont {Gegenwart}}, \ and\ \bibinfo {author} {\bibfnamefont
  {R.}~\bibnamefont {Valent{\'{\i}}}},\ }\href {\doibase
  10.1088/1361-648x/aa8cf5} {\bibfield  {journal} {\bibinfo  {journal} {J.
  Phys.:Condens. Matter.}\ }\textbf {\bibinfo {volume} {29}},\ \bibinfo {pages}
  {493002} (\bibinfo {year} {2017})}\BibitemShut {NoStop}%
\bibitem [{\citenamefont {Hermanns}\ \emph {et~al.}(2018)\citenamefont
  {Hermanns}, \citenamefont {Kimchi},\ and\ \citenamefont
  {Knolle}}]{KitaevMaterial-3}%
  \BibitemOpen
  \bibfield  {author} {\bibinfo {author} {\bibfnamefont {M.}~\bibnamefont
  {Hermanns}}, \bibinfo {author} {\bibfnamefont {I.}~\bibnamefont {Kimchi}}, \
  and\ \bibinfo {author} {\bibfnamefont {J.}~\bibnamefont {Knolle}},\ }\href
  {\doibase 10.1146/annurev-conmatphys-033117-053934} {\bibfield  {journal}
  {\bibinfo  {journal} {Annu. Rev. Condens. Matter Phys.}\ }\textbf {\bibinfo
  {volume} {9}},\ \bibinfo {pages} {17} (\bibinfo {year} {2018})}\BibitemShut
  {NoStop}%
\bibitem [{\citenamefont {Knolle}\ and\ \citenamefont
  {Moessner}(2019)}]{KitaevMaterial-4}%
  \BibitemOpen
  \bibfield  {author} {\bibinfo {author} {\bibfnamefont {J.}~\bibnamefont
  {Knolle}}\ and\ \bibinfo {author} {\bibfnamefont {R.}~\bibnamefont
  {Moessner}},\ }\href {\doibase 10.1146/annurev-conmatphys-031218-013401}
  {\bibfield  {journal} {\bibinfo  {journal} {Annu. Rev. Condens. Matter
  Phys.}\ }\textbf {\bibinfo {volume} {10}},\ \bibinfo {pages} {451} (\bibinfo
  {year} {2019})}\BibitemShut {NoStop}%
\bibitem [{\citenamefont {Takagi}\ \emph {et~al.}(2019)\citenamefont {Takagi},
  \citenamefont {Takayama}, \citenamefont {Jackeli}, \citenamefont
  {Khaliullin},\ and\ \citenamefont {Nagler}}]{KitaevMaterial-5}%
  \BibitemOpen
  \bibfield  {author} {\bibinfo {author} {\bibfnamefont {H.}~\bibnamefont
  {Takagi}}, \bibinfo {author} {\bibfnamefont {T.}~\bibnamefont {Takayama}},
  \bibinfo {author} {\bibfnamefont {G.}~\bibnamefont {Jackeli}}, \bibinfo
  {author} {\bibfnamefont {G.}~\bibnamefont {Khaliullin}}, \ and\ \bibinfo
  {author} {\bibfnamefont {S.~E.}\ \bibnamefont {Nagler}},\ }\href {\doibase
  10.1038/s42254-019-0038-2} {\bibfield  {journal} {\bibinfo  {journal} {Nat.
  Rev. Phys.}\ }\textbf {\bibinfo {volume} {1}},\ \bibinfo {pages} {264}
  (\bibinfo {year} {2019})}\BibitemShut {NoStop}%
\bibitem [{\citenamefont {Janssen}\ and\ \citenamefont
  {Vojta}(2019)}]{KitaevMaterial-6}%
  \BibitemOpen
  \bibfield  {author} {\bibinfo {author} {\bibfnamefont {L.}~\bibnamefont
  {Janssen}}\ and\ \bibinfo {author} {\bibfnamefont {M.}~\bibnamefont
  {Vojta}},\ }\href {\doibase 10.1088/1361-648x/ab283e} {\bibfield  {journal}
  {\bibinfo  {journal} {J. Phys.: Condens. Matter.}\ }\textbf {\bibinfo
  {volume} {31}},\ \bibinfo {pages} {423002} (\bibinfo {year}
  {2019})}\BibitemShut {NoStop}%
\bibitem [{\citenamefont {Koga}\ \emph {et~al.}(shed)\citenamefont {Koga},
  \citenamefont {Murakami},\ and\ \citenamefont {Nasu}}]{KogaArxiv}%
  \BibitemOpen
  \bibfield  {author} {\bibinfo {author} {\bibfnamefont {A.}~\bibnamefont
  {Koga}}, \bibinfo {author} {\bibfnamefont {Y.}~\bibnamefont {Murakami}}, \
  and\ \bibinfo {author} {\bibfnamefont {J.}~\bibnamefont {Nasu}},\ }\href@noop
  {} {\bibfield  {journal} {\bibinfo  {journal} {preprint}\ ,\ \bibinfo {pages}
  {arXiv:2104.02182}} (\bibinfo {year} {unpublished})}\BibitemShut {NoStop}%
\bibitem [{\citenamefont {Arakawa}\ and\ \citenamefont
  {Yonemitsu}(2021)}]{Arakawa}%
  \BibitemOpen
  \bibfield  {author} {\bibinfo {author} {\bibfnamefont {N.}~\bibnamefont
  {Arakawa}}\ and\ \bibinfo {author} {\bibfnamefont {K.}~\bibnamefont
  {Yonemitsu}},\ }\href {\doibase 10.1103/PhysRevB.103.L100408} {\bibfield
  {journal} {\bibinfo  {journal} {Phys. Rev. B}\ }\textbf {\bibinfo {volume}
  {103}},\ \bibinfo {pages} {L100408} (\bibinfo {year} {2021})}\BibitemShut
  {NoStop}%
\end{thebibliography}%

\end{document}